\documentclass[preprint,11pt]{elsarticle}

\usepackage{amssymb}
\usepackage{pdfpages}

\usepackage{hyperref}

\journal{xxx}

\begin{document}

\begin{frontmatter}

\title{An approach based on Open Research Knowledge Graph for Knowledge Acquisition from scientific papers}

\author[uy1]{Azanzi Jiomekong}

\address[uy1]{University of Yaounde I, Faculty of Sciences, Cameroon \\ fidel.jiomekong@facsciences-uy1.cm}

\author[uat]{Sanju Tiwari}

\address[uat]{Universidad Autonoma de Tamaulipas, Mexico, India \\ tiwarisanju18@ieee.org}

\begin{abstract}
A scientific paper can be divided into two major constructs which are Metadata and Full-body text. Metadata provides a brief overview of the paper while the Full-body text contains key-insights that can be valuable to fellow researchers. To retrieve metadata and key-insights from scientific papers, knowledge acquisition is a central activity. It consists of gathering, analyzing and organizing knowledge embedded in scientific papers in such a way that it can be used and reused whenever needed. Given the wealth of scientific literature, manual knowledge acquisition is a cumbersome task. Thus, computer-assisted and (semi-)automatic strategies are generally adopted. Our purpose in this research was two fold: curate Open Research Knowledge Graph (ORKG) with papers related to ontology learning and define an approach using ORKG as a computer-assisted tool to organize key-insights extracted from research papers. This approach was used to document the "epidemiological surveillance systems design and implementation" research problem and to prepare the related work of this paper. It is currently used to document "food information engineering", "Tabular data to Knowledge Graph Matching" and "Question Answering" research problems and "Neuro-symbolic AI" domain.
\end{abstract}


\end{frontmatter}


\section{Introduction}
\label{introduction}
Scientific papers are one of the greatest assets for scientists. They constitute one of the primary source of knowledge for researchers, and sometimes for decision makers \cite{IEResearchPapersJayaram2017,IEArticleParantu2003}. They are recorded, indexed and disseminated in scientific publication repositories such as ISI Web of Knowledge, IEE Xplore, Springer, ACM, ScienceDirect, Scopus, Semantic Scholar, etc. In consequence, the body of scientific literature is growing at an enormous rate \cite{IEArticleParantu2003,ImprovingAccessAuer2020,surveyIEScienPaperZara2018,KEReviewAbdul2020}. This wealth of scientific knowledge is widely disseminated to users who now possess an unprecedented problem of access to scientific literature \cite{surveyIEScienPaperZara2018,KEReviewAbdul2020,LimitIETechKiran2019}. In effect, this increase in scientific content poses significant challenges for the researchers who want to sort through, read, understand, compare, and build upon to determine for instance, the state of art in their respective field of interest \cite{surveyIEScienPaperZara2018,IEGISPapersAcheson2021}.

Globally, a scientific paper can be divided into two major constructs which are Metadata and Full-body text \cite{surveyIEScienPaperZara2018,KEReviewAbdul2020}. Metadata provides a brief overview of the scientific papers and the Full-body text contains valuable information that is beneficial to fellow researchers. To retrieve metadata and key-insights from scientific papers, Knowledge Acquisition (KA) \cite{KAGuideHayes1996} is a central activity in research.

Knowledge are facts, information and skills acquired through experience or education for understanding of a subject area \cite{Djuric2005}. Concerning scientific papers, knowledge are metadata provided by editors and authors, and key-insights provided by authors which are used by fellow researchers to understand the scientific paper content. Knowledge Acquisition from scientific papers refers to the method for gathering, analyzing and organizing knowledge embedded in these papers. This involves the extraction of structured content in the form of entities, relations, facts, terms, and other types of information that may help researchers to understand the papers and get insights from them \cite{LimitIETechKiran2019}. After its acquisition, knowledge is organized in such a way that it can be used and reused whenever needed. Globally, knowledge acquisition can happen through a wide variety of strategies that vary from completely manual to totally automated \cite{IEResearchPapersJayaram2017,IEGISPapersAcheson2021,KAGuideHayes1996}. Concerning knowledge acquisition from scientific papers, we distinguished the manual process \cite{IEResearchPapersJayaram2017,IEGISPapersAcheson2021,DiscoveringEcologicallyKarl2013,AmbiguousGeographiesJared2016} and the (semi-)automatic process.

Given the amount of scientific papers that a domain may have, the manual process can be a cumbersome job, time consuming, not scalable and not efficient. To reduce the burden of KA, computer-assisted and (semi-)automated strategies are proposed \cite{IEArticleParantu2003} for processing and cataloging scientific knowledge, for assisting researchers to choose their papers, navigate amongst papers, compare them and get insights from them.

During the last decades, many researchers have contributed to the automatic extraction of metadata from scientific papers. Multiple rule-based, machine learning and NLP techniques have been proposed \cite{IEResearchPapersJayaram2017,surveyIEScienPaperZara2018,KEReviewAbdul2020}. Concerning knowledge extraction from the full-body text, it has been reported that key-insights are deeply hidden in the text and are difficult to extract \cite{IEResearchPapersJayaram2017,IEArticleParantu2003,ImprovingAccessAuer2020,surveyIEScienPaperZara2018,KAMagdi2007}. To allow researchers to collaboratively build the body of knowledge from their domain and research interest, we propose a computer-assisted knowledge acquisition approach. This is based on the use of Open Research Knowledge Graph (ORKG) \cite{ImprovingAccessAuer2020} for automatic acquisition of metadata and manual annotation of the paper with key-insights to produce a semantic description of the scientific knowledge of the domain in a Knowledge Graph (KG). Once extracted and organized, research contributions can be compared using annotated tables and graphics.

This approach is inspired by the use of ORKG in our research since three years to: (1) Organize and compare research contributions so as to build a large dataset of up-to-date knowledge for the following research problems: "ontology learning", "epidemiological surveillance systems design and implementation",  "food information engineering", "Tabular data to Knowledge Graph Matching", "Question Answering", and "information extraction from scientific papers'', and "Neuro-symbolic AI" domain. (2) Organize research so as to facilitate the update and improvement with the contributions of fellow researchers working on the same research problem or the same domain.

In the rest of the paper, we present Open Research Knowledge Graph in Section \ref{orkg} and the research methodology in Section \ref{orkgCuration}. In Section \ref{orkgApproach}, we present the approach we propose for Knowledge Acquisition from scientific papers using ORKG and in Section \ref{orkgApproach} we present the use of this approach on 3 use cases: "epidemiological surveillance systems'', "food information engineering" and "knowledge extraction from scientific papers''. The latter use case was used to write the related work of this paper (Section \ref{relatedWork}). Finally, in Section \ref{conclusion}, we conclude.

\section{Scientific papers description}
\label{scientificPapers}
On the basis of its structure, knowledge contained in a scientific paper is broadly classified into two major categories which are metadata (see Section \ref{scientificPapers:metadata}) and key-insights (see Section \ref{scientificPapers:keyInsight}).

\subsection{Metadata}
\label{scientificPapers:metadata}
Metadata information is used either for scientific paper recommendation by research repositories, or to furnish a brief overview about a scientific paper. The latter allows a researcher to decide the paper's relevance with their domain of interest \cite{surveyIEScienPaperZara2018}. Metadata can be defined in two main components: those that are assigned by the authors (such as the title of the paper, Abstract, Keywords, etc.) and those that are assigned by the editors (such as BibTex and/or DOI, Copyright, Date of publication, etc.).

\paragraph{Metadata extraction}
Metadata extraction (ME) refers to the identification and extraction of metadata elements. In order to perform ME, there exist multiple datasets that vary on the basis of article's sources, publication venues, etc. On these datasets, multiple automatic approaches are applied. They use the DOI, BibTex or the title of the paper to search and fetch these papers from scientific repositories. Thereafter, rule and/or Machine Learning techniques are used to extract these metadata \cite{surveyIEScienPaperZara2018}.

\subsection{Key-insights}
\label{scientificPapers:keyInsight}
The full body text of the paper hides the key-insights/knowledge that the readers need to extract in order to understand the paper. Even if the authors can choose their own way to organize the full-body text, the journal's Guide for Authors provides to the authors a template composed of the different sections that the paper may include. Whatever the organization of the paper provided by the authors, one can identify the introduction, Research methods and methodologies, Results, Discussion, Related work and/or literature review and conclusion.

\paragraph{Knowledge extraction from the full-content of a scientific paper}
From the full-body text of a scientific paper, entities such as research domain, research problem, methodology of the research, methods, models, algorithms, processes, data-source, data-sets, tools, evaluation measures, results achieved, limitations of the research, future directions, etc. are extracted by the readers in order to understand the paper. These entities once extracted can be organized into instances. These instances can be grouped into classes with associated properties.

From classes, the following relations can be extracted:
\begin{itemize}
	\item \textbf{Taxonomy:} this relation organizes classes into a hierarchical relation. For instance, we used it to organize ontology learning research problems using a taxonomy of research problems related to ontology learning. This taxonomy shows that ontology learning research can be divided into the following research problems: "Ontology learning from unstructured sources", "Ontology learning from semi-structured sources", and "ontology learning from structured sources". These research problems can also be divided by considering different data sources.
    
	\item \textbf{Association:} This is the link used to define that two classes are related to each other. For example, in the sentence: "Jiomekong et al. proposed to use Hidden Markov Models to extract knowledge from source code", we identified the classes "Techniques" and "Knowledge source". So, a relation named "extract" of application between the class "Technique" and the class "Knowledge source" can be established. The instance of the class "Technique" is then "Hidden Markov Models" and the instance of the class "Knowledge source" is "Source code" and we can have the following statement: "Hidden Markov Models are used to extract knowledge from source code".
\end{itemize}
Once extracted, key-insights are grouped into research contributions and used to write state-of-the-art. In the latter, many tables and graphics are used to compare research contributions of several authors.

The extraction of key-insights from scientific papers are generally manual. However, knowledge extracted are sparse in different data sources (scientific papers, research computers, etc.), with the risk of being forgotten, lost and making it difficult to compare background research problems to up-to-date ones. In the next section, we present how Open Research Knowledge Graph can be used as a computer-assisted tool to solve these problems.

\section{Open Research Knowledge Graph}
\label{orkg}
In this section, we present an overview of ORKG (Section \ref{orkgOverview}) and the main features used during this research (Section \ref{orkgFeatures}).

\subsection{Overview of ORKG}
\label{orkgOverview}
ORKG is an open research infrastructure designed to acquire, publish and process structured scholarly knowledge published in the scholarly literature \cite{ImprovingAccessAuer2020,ComparingReConAllard2019}. It is built according to the principles of Open Science, Open Data, and Open Source.
\begin{itemize}
	\item \textbf{Open Science:} ORKG resources such as comparisons of scientific papers and the smart reviews can be developed through collaborative network of researchers. Once published, these resources are freely available to anyone who wants to learn about the research question and/or to contribute.
    
	\item \textbf{Open Data:} All the data ingested in ORKG is in a machine readable format and open to everyone who needs to share, use, re-use, modify, and share the modified version. The only restriction concerns contributing to an ORKG resource. This restriction consists of having an ORKG account.
    
	\item \textbf{Open Source:} the source code of ORKG is available to the general public.
\end{itemize}

Thus, all the ORKG source code, information, data are available under open licenses \cite{ImprovingAccessAuer2020}. To date, ORKG indexes more than 10,000 research papers corresponding to more than 5000 research problems (corresponding to 1237 research fields), more 1000 comparisons, 224 templates, 1000 users, 2216 benchmarks\footnote{\url{https://www.orkg.org/orkg/stats}}.

\subsection{ORKG features}
\label{orkgFeatures}
To help researchers structure and organize the research contributions extracted from scientific papers, ORKG provides a set of features. In this Section, we present the ones we used during our research.

\paragraph{Add research problems} The research problems of a research area can be described independently, provided with relevant sources and assigned to a taxonomy of research problems \cite{ImprovingAccessAuer2020}. For instance, with ORKG, we can define a taxonomy of research problems related to ontology learning.

\paragraph{Add papers} ORKG represents an article with \cite{ImprovingAccessAuer2020}:
\begin{enumerate}
	\item \textbf{Article metadata:} The article metadata involves the bibliographic information such as article title, authors, journal, book title, etc. ;
	\item \textbf{Semantic description of the article:} These are key-insights of the papers extracted and annotated by researchers by following the Subject-Predicate-Object triple principle.
\end{enumerate}

\begin{figure}
	\centering
	\includegraphics[scale=0.35, angle=90]{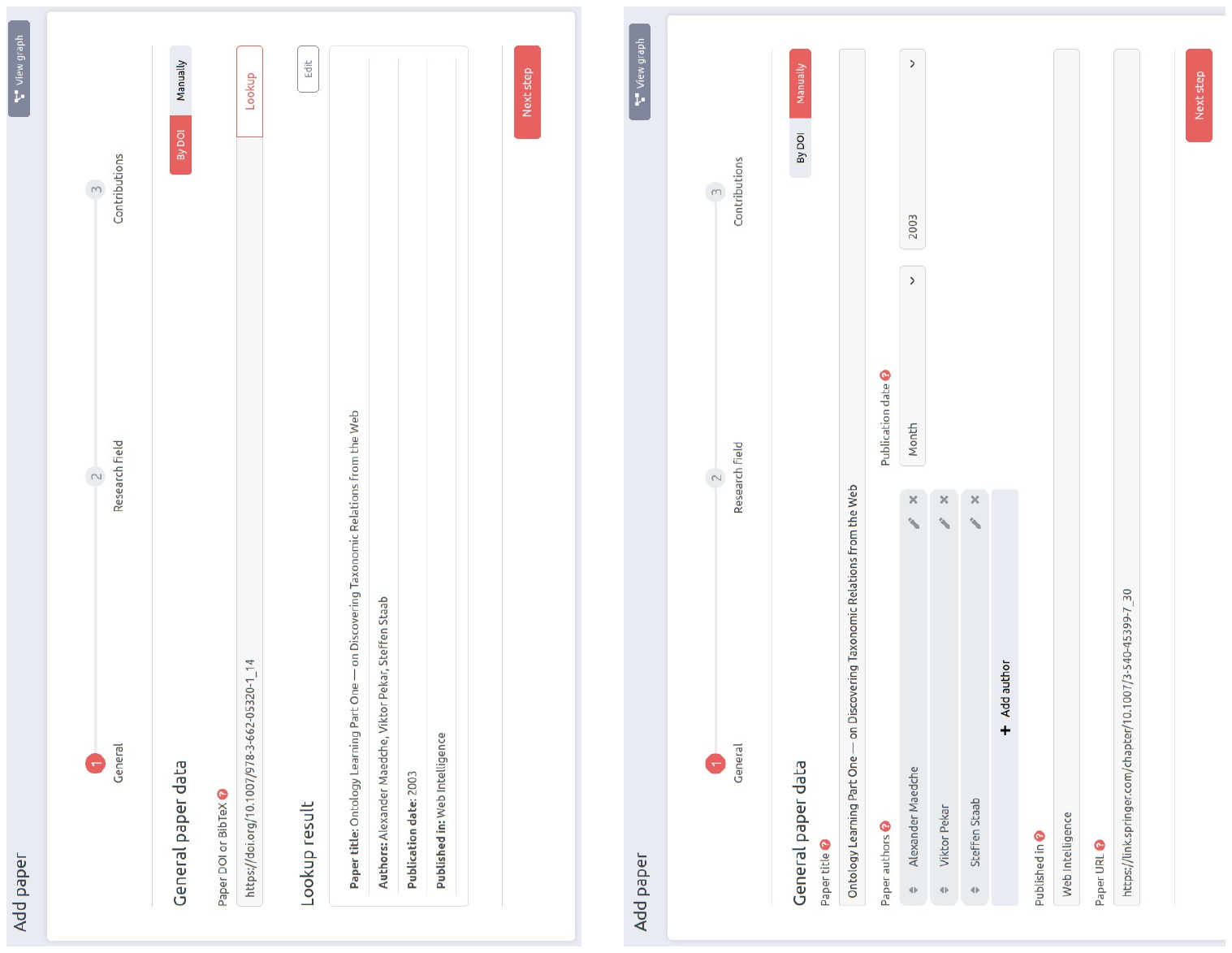}
	\caption{Manual (first picture) and automatic (second picture) acquisition of meta-data of a paper}
	\label{fig:addPaperWizard}
\end{figure}

The article metadata and its semantic description are used to annotate the paper. To this end, researchers are allowed to add papers manually or (semi-)automatically to ORKG  (see Fig. \ref{fig:addPaperWizard}) \cite{ImprovingAccessAuer2020}:
\begin{itemize}
	\item During the manual process, all the key-metadata (title, author, etc.) and key-insights (research domain, research problem, research tools, etc.) of the papers are manually acquired by the researchers and added to ORKG using a wizard provided by the system.
    
	\item To semi-automatically add an article to the system, the key-metadata of the article such as the paper title, DOI or BibTex are provided to the ORKG wizard. These informations are used by the system to fetch the articles key-metadata. Once extracted, these informations are presented to the users so that they can complete missing metadata. Once the metadata are added to the paper, the researchers use a wizard provided by ORKG to semantically describe the paper with key-insights they extracted manually.
\end{itemize}

\begin{figure*}[t]
	\centering
	\includegraphics[scale=0.3]{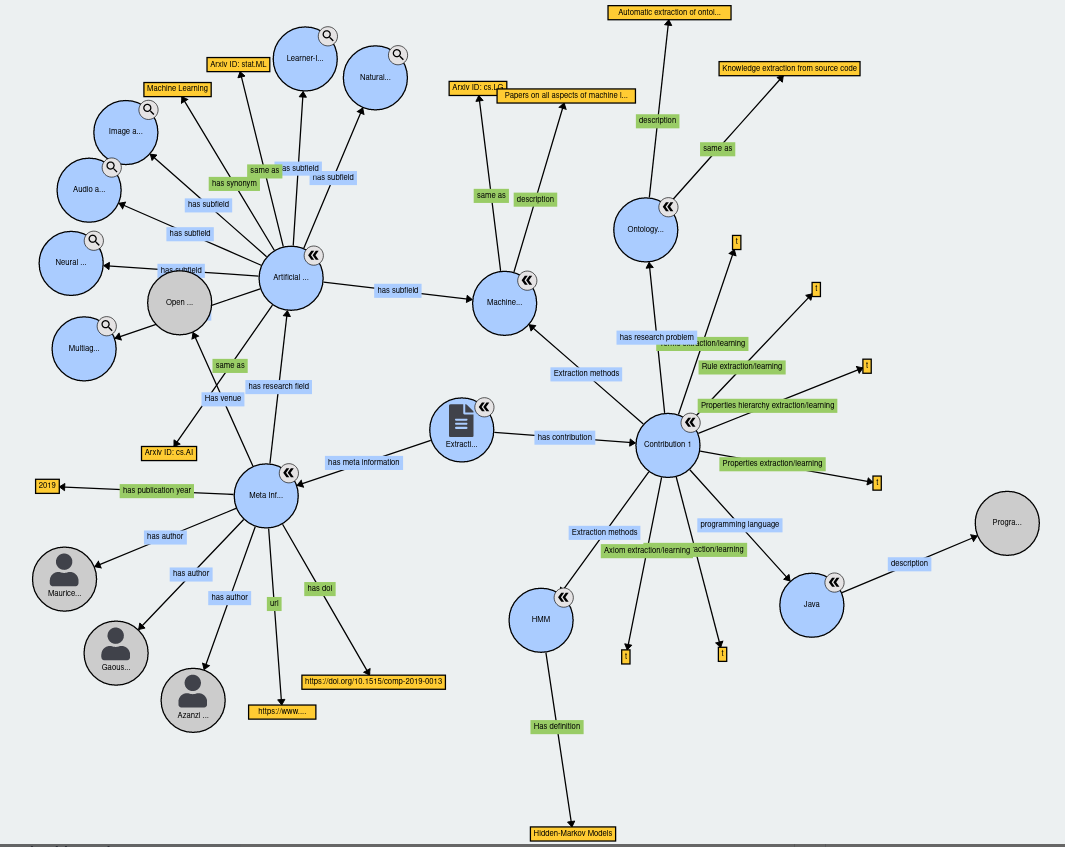}
	\caption{RDF graph representation of a paper with their metadata and contributions}
	\label{fig:rdfPaperGraph}
\end{figure*}

\paragraph{Semantic description of research papers} The semantic description of research papers consists of the annotation of these papers with key-insights extracted from them and to organize these elements into research contributions. This allows us to put the paper in machine readable form following the RDF subject-predicate-object paradigm. The ORKG annotation feature is a flexible system that allows users to reuse existing predicates and functions or to create and add their own predicates (properties or attributes). The description of the entities in human readable form allows researchers to have a common understanding of the data between the various stakeholders. Fig. \ref{fig:rdfPaperGraph} presents an example of a graph representation of a paper with their metadata and key-insights organized in paper contributions.

The graph of Fig. \ref{fig:rdfPaperGraph} presents the paper entitled "Knowledge extraction from java source code using Hidden-Markov Models". The key-insights extracted are:
\begin{itemize}
    \item The research problem which is "Knowledge extraction from source code",
    \item Different types of knowledge that are extracted,
    \item Techniques that are used during the extraction process,
    \item The programming language from which the source code was written.
\end{itemize}

\begin{figure*}
	\centering
	\includegraphics[scale=0.2]{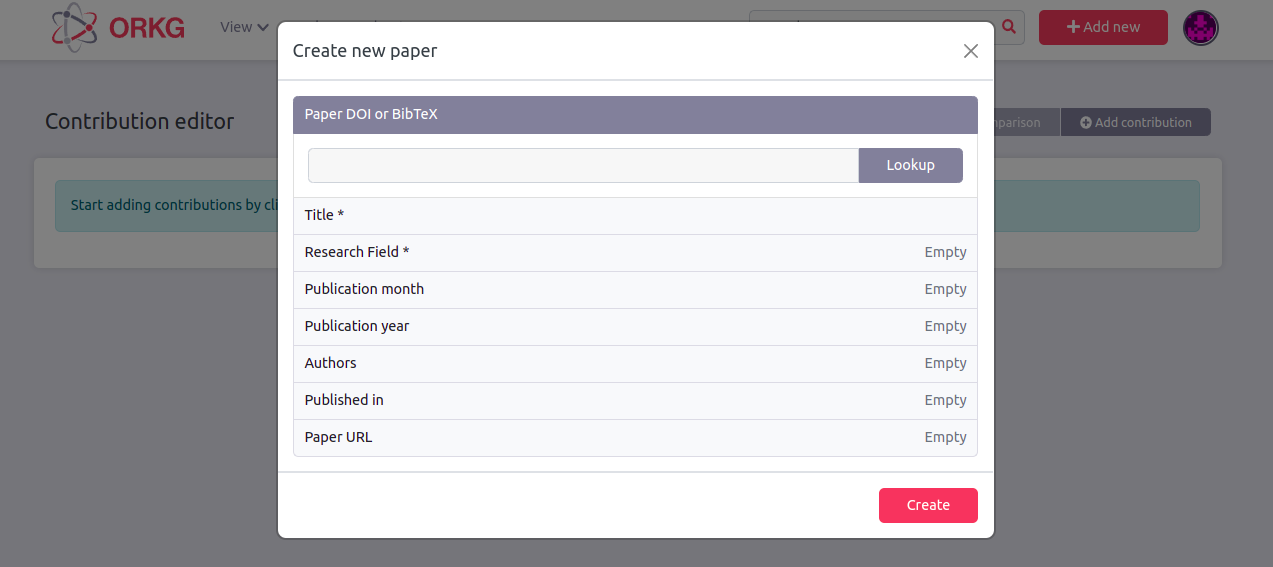}
	\caption{Adding a paper using the comparison table wizard}
	\label{fig:comparEditAddPaper}
\end{figure*}

\paragraph{Add research contributions to a paper}
In ORKG, each paper consists of at least one research contribution which addresses at least one research problem and is further described with contribution data including materials, methods, implementation, results or other key-insights. The paper of the Fig. \ref{fig:rdfPaperGraph} presents one research contribution. These contributions can be compared between them or by other contributions from other papers \cite{ComparingReConAllard2019} in an ORKG comparison table. Papers can be added to ORKG during the creation of the comparison table as presented by the Fig. \ref{fig:comparEditAddPaper}.

We consider in this research that all the key-insights in the research paper such as the definition of research problem, materials and methods used, results obtained, lessons learned, etc. are grouped into research contributions. During the adding paper process, a default research contribution containing the key-insights such as research domain and research problem is filled by the user. Research contributions are described in a structured and semantic way as a Knowledge Graph (see Fig. \ref{fig:rdfPaperGraph}). Therefore, the information will not be only readable by humans, but also by machines \cite{ImprovingAccessAuer2020}.

\begin{figure*}[t]
	\centering
	\includegraphics[scale=0.2]{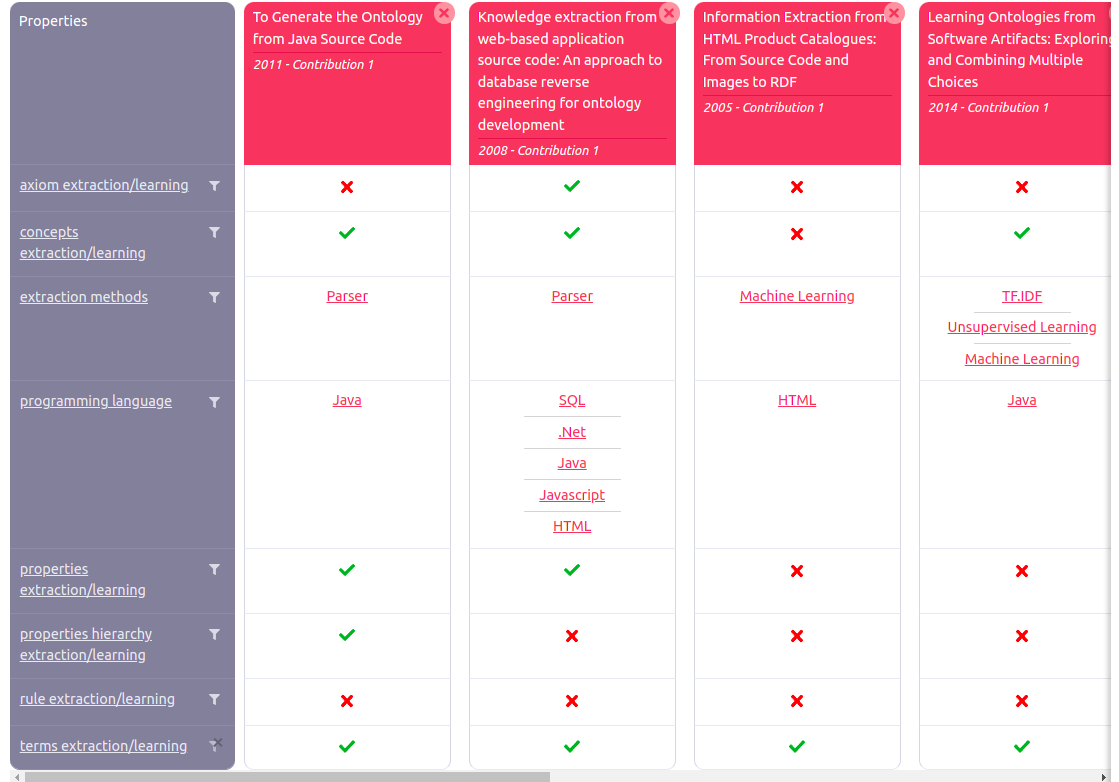}
	\caption{A table presenting the comparison of research contributions of papers related to ontology learning from source code}
	\label{fig:comparisonTable}
\end{figure*}

\paragraph{Comparing research papers} The structured content descriptions of scientific contributions presented above are presented in such a way that the contribution becomes comparable with other articles of the research domain. Therefore, the structured semantic representation of scientific knowledge in the knowledge graph makes it possible to automatically create literature comparisons. Allard et al. \cite{ComparingReConAllard2019} present a workflow designed to compare research contributions in ORKG. Fig. \ref{fig:comparisonTable} presents an example of a comparison table built using this workflow. This is the comparison of research contributions of papers related to ontology learning from source code. The comparison table can be published with DOI, exported in various formats such as RDF, LaTeX, PDF, CSV and integrated in a literature review. The comparison table link can be shared to other researchers, so that they can improve the comparison by correcting errors or adding missing information.

\paragraph{Templates} Scientific papers usually lack a formal metrical structure. It comprises full grammatical sentences, paragraphs in which key-insights are hidden. Identifying and structuring research contributions found in scientific papers is not always an easy task for a research student or newcomers in the domain. This is because the description of scientific findings is complex and is based on expert knowledge. On the other hand, the researcher should decide in which granularity a research contribution should be described so as to be comparable.

The goal of the template is to highlight for a research problem, a set of key-insights that may be found in a scientific paper addressing this research problem. It specifies the structure of scientific information so that \cite{ImprovingAccessAuer2020}: (1) Fellow researchers can compete with more key-insights, (2) New researchers can rapidly get insights in the research domain.

Templates can then be reused in the description of research contributions to facilitate data entry and ensure comparability. For instance, we built a template for documenting existing datasets for metadata extraction from scientific papers \footnote{\url{https://www.orkg.org/orkg/template/R277000}}.

\paragraph{Graph visualization} Once added to a paper, the graph representing the research contribution is generated. This graph can be used for the exploration of scientific contribution.

\begin{figure*}
	\centering
	\includegraphics[scale=0.4, angle=90]{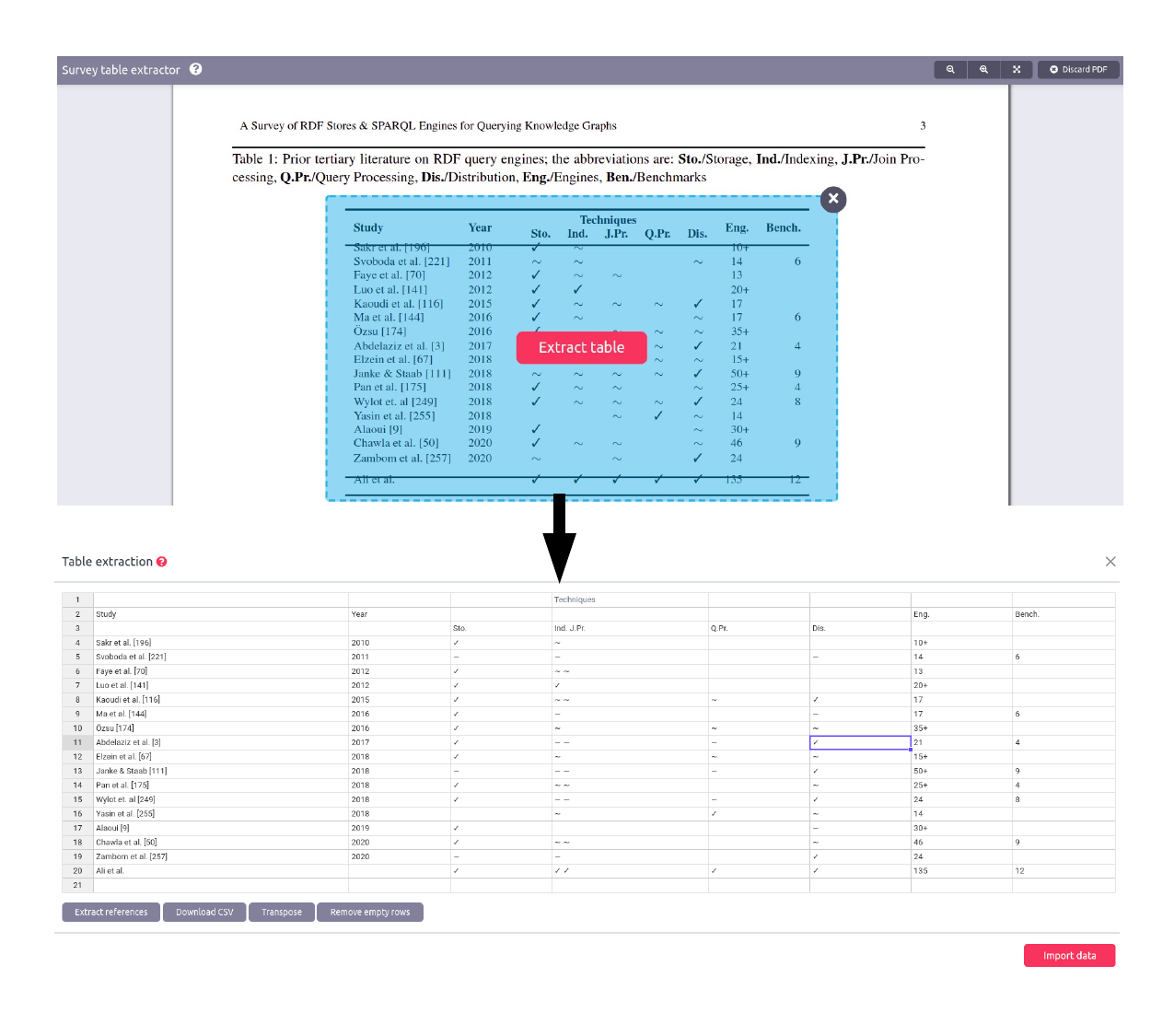}
	\caption{Extracting key-insights on graph databases using ORKG}
	\label{fig:extractSurvey}
\end{figure*}

\paragraph{Importing survey papers: } Survey articles present an overview of the state-of-the-art for a specific area. Within survey articles, some overviews or summaries are often presented in (semi-)structured tabular format. From these tables, information on key-insights of the papers involved in the literature review can be extracted (semi-)automatically as follow: the first step consists of extracting the key-metadata and the key-insights from the table and building a comparison table; the second step involves fixing potential extraction errors and adding additional metadata or key-insights that was not automatically extracted. The Fig. \ref{fig:extractSurvey} presents the extraction wizard.

\paragraph{Smart Review} After the creation of a comparison, a researcher may create a smart review for giving an overview on research addressing a particular research question. To this end, ORKG furnishes a "What You See Is What You Get (WYSIWIG)" editor allowing researchers to create a structured overview of the literature.

\paragraph{Collaborative work on literature review} In ORKG, collaborative work allows a whole community of researchers to collaboratively build the state of the art of a research problem. In effect, many authors working on the same research problems can gather to add and modify research contributions of a scientific paper. Once these contributions are compared using ORKG comparison tables and used to write smart reviews they can be shared amongst other researchers in order to get their viewpoints. To this end, contributions and smart reviews are versioned so that all changes can be discussed by the professional community, updated and new versions published. If new literature is published, it is easy to continuously expand the comparison, which thus continues to reflect the current state of knowledge in a comparable way.

\section{Research Methodology}
\label{orkgCuration}
The research methodology consists of action research. Action research methodology is used when major challenges cannot be studied without implementing them, and where implementation implies a long term commitment because effect may take time to emerge \cite{empiricalStandSoftEnRalph2020}. In our case, we wanted to explore, test and evaluate the different features that can be used for knowledge acquisition from scientific papers using Open Research Knowledge Graph as a computer assistant tool. Given that action research allows us to plan, implement, revise, then implement, lending itself to an ongoing process of reflection and revision, we thought it necessary to use this research methodology.

Globally, the research methodology consisted of a set of aggregated interventions to curate the papers. These interventions involved a series of actions taken during the curation of scientific papers. At its end, we come up with a research methodology that is reported in this section. This research methodology consists of the Pre-Intervention (Section \ref{researchMethodology:preIntervention}) and the Intervention (Section \ref{researchMethodology:intervention}) phase of the Action research methodology. The Post-Intervention presented in Section \ref{postIntervention} consists of the use of the methodology presented in this section in three use cases.

\subsection{Pre-Intervention}
\label{researchMethodology:preIntervention}
During the ORKG curation, we particularly worked in the domain of Semantic Web. The Pre-Intervention step consists of the definition of the research objective and the organization of the curation.

\subsubsection{Research objective}
We started this research in 2021 with the objective to document the research problem: "ontology learning". In effect, ontology learning is the automatic or semi-automatic extraction of ontological knowledge from unstructured, semi-structured or fully structured knowledge sources in order to build an ontology from them  with little human intervention \cite{AzanziCT2019,KONYS20182194,ShamsfardMehrnoush2003,zhou2007ontology}. This choice was motivated with our recent work on ontology learning from source code \cite{AzanziCT2019}. This is the automatic extraction of ontological knowledge from software source code. Thus, we decided to curate this paper first. Thereafter, we curated the related work of this paper.

\subsubsection{Selection of papers to curate}
We started with the selection of the paper we wrote on ontology learning from source code \cite{AzanziCT2019}. Thereafter, we selected all the papers related to ontology learning from other data sources that were cited in this paper. For the "ontology learning from datasources" that were not cited such as "ontology learning from folksonomies" or "ontology learning from thesaurus", we used the famous research repository Semantic Scholar to search for relevant papers. The keyword "ontology learning from xxx" (where xxx represents the data source) were entered in the search bar of the Semantic Scholar platform. We used the papers titles, short abstract provided by Semantic Scholar and paper abstract provided by the authors to select relevant papers. Given that our goal was mainly to curate some papers and understand how to use ORKG for knowledge acquisition from scientific papers, we only choose the papers on the first page results. Key-insights were extracted from these papers, comparison metrics defined and used to compare these papers.

\subsubsection{Work organization}
Globally, the curation of ORKG involves two groups of people: the ORKG team and the curators. The ORKG team is a group of persons responsible for the organization of the curation meetings, description of tasks of curators, training of curators on the use of the tool and support when they have any difficulties. Before we start the curation in June 2021, a training session was made by the ORKG team. This session was oriented on the presentation of ORKG features, the creation of comparisons using the ORKG comparison editor. During the period of curation, many demos on the creation of comparisons, templates, and smart reviews were made.

To support the curators and respond to all their difficulties, a Mail and a Skype Group were created and a bi-monthly meeting was set-up. During these meetings, we were having 5-10 minutes time to present our work: adding papers, creating comparisons tables, templates, smart review, etc. Thereafter, the questions and the remarks were posed in order to help to ameliorate the work. The meetings were recorded with Skype so that we can watch it later. During these meetings, the comparisons of papers made by the curators were discussed so that they can update and correct errors. Examples of discussions concern the definition of classes, properties, the coding of knowledge extracted from the scientific papers, etc.

\subsection{Intervention}
\label{researchMethodology:intervention}
During the intervention phase, we extracted key-insights from scientific papers and we used these key-insights to create comparison criteria (these are ORKG properties). Thereafter, these comparison criteria were used to compare these scientific papers using the ORKG comparison table. This is an iterative and incremental process during which the experience we got during the creation of one comparison table is used to ameliorate this comparison table and create the new ones. Comparison tables were evaluated by the ORKG team and fellow researchers and refined. For instance, the first comparison\footnote{\url{https://www.orkg.org/orkg/comparison/R138057}} were refined until it was accepted as well organized by the ORKG team and some colleagues working in the domain of ontology learning. Globally, papers related to the following thematic were curated:
\begin{itemize}
	\item Ontology learning from Thesaurus (5 papers),
	\item Ontology learning from Glossaries (2 papers),
	\item Ontology learning from taxonomies (2 papers),
	\item Ontology learning from XML (15 papers),
	\item Ontology learning from UML (4 papers),
	\item Ontology learning from source code (9 papers)
	\item Ontology learning from folksonomies (6 papers)
	\item Ontology learning from images (2 papers)
	\item Ontology Learning from Entity Relation Model (9 papers).
\end{itemize}
At the end, 54 papers were curated, 9 comparison tables were created using these papers, and one smart review on ontology learning from images. In the following paragraphs, we present how we proceed to create these comparisons, the lessons learned and main finding that was used to ameliorate our work.

\subsubsection{Creation of the first comparison}
The first work we did was to create the first comparison of papers. Nine papers related to "ontology learning from source code" research problem were read, knowledge extracted and ingested into the ORKG platform. To this end, we firstly created a comparison table and using the ORKG comparison table wizard, we added papers to ORKG. These papers were added manually and (semi-)automatically to ORKG:

\begin{itemize}
	\item We used the manual process for the papers that do not have DOI or BibTex. During this process, all the key-metadata (title, author, etc.) and key-insights (research domain, research problem, etc.) of the papers are manually acquired and added to ORKG using a wizard provided by the system.
    
	\item To (semi-)automatically add an article to ORKG, we used the DOI or BibTex to automatically fetch the articles metadata. Once extracted, missing informations are completed and the paper is annotated with key-insights extracted manually.
\end{itemize}
Once a paper is added, a graph representing the research contribution allows us to visualize and verify that the information on the paper is well structured.

The comparison table of ontology learning papers from source code contains the following elements:
\begin{itemize}
	\item The first column of the table contains properties, which can also be seen as a comparison criteria.
	\item The rest of the column corresponds to papers that are compared.
	\item For each row, the corresponding insight extracted from the paper is presented, so that these elements can be used to compare papers together.
\end{itemize}
From this comparison, we learned how to organize research contributions using ORKG. The exchange with the ORKG team and some colleagues working in the domain of knowledge engineering allowed us to ameliorate this comparison and a new version was published. We found the tool interesting to save our work so to reuse later in scientific papers as additional material or related work. This motivates us to create more comparisons and explore the other features of the system.

\subsubsection{Creation of other comparisons}
The creation of the first comparison allowed us to master the use of the comparison wizard. Therefore, 7 more comparisons were created. These comparisons gave rise to a refinement iteration in order to identify all potential knowledge that will be converted into classes, relations and properties and that will be used to build a high-quality and comparable structured scientific knowledge for "ontology learning" research problems. The aim of this structure being to create a common Semantic Model to reflect contributions to “ontology learning” research problems. For instance, for ontology learning methods such as "TF.IDF", "Unsupervised Learning", "deep learning", "Neural Network", we decided to group them and to create a class labeled "Learning method".

\paragraph{Lesson learned}
The comparisons presented above led to the following lessons:
\begin{itemize}
	\item Structure and describe research contribution is not an easy task: During the creation of comparisons presented above, we learned that to structure and describe a research paper is not an easy task. In effect, describing research contributions and making them comparable is complex because the granularity of comparison should be decided. For instance, should we consider the comparison of methods for knowledge extraction from "unstructured sources" and "structured sources" or should we go further and compare unstructured data sources such as "text", "images", with the structured ones such as "databases", "UML models"? Given that we wanted fellow researchers to see the methodologies, methods and tools for ontological knowledge extraction from knowledge sources, we decided to add a property that indicate if the data source is unstructured and the type of the data source (e.g., "text", database", etc.)
	\item Find the accurate property for the comparison is not an easy task: It is recommended to reused as much as possible existing ORKG properties that were created by other researchers. However, we found it difficult because one has to scroll down any time one wants to add a property to a contribution (time consuming). On the other hand, after some time, the description of a property can be forgotten or unknown (for those who did not input them). This makes it difficult to find the right property to use in the comparison tables. Fortunately, the ORKG wizard provides the properties description. However, many properties had the same name and no description.
\end{itemize}

\paragraph{Insight} To solve the above problems, we found it necessary to use the ORKG template feature to structure scientific papers related to "ontology learning". This template is supposed to contain all the properties that should be compared. To facilitate its accessibility, we decided to add descriptions to all the properties used. Thus, to add a contribution from a paper related to the "ontology learning" research problem, this template is used. This template is a standardized tool that can be refined and used to compare as many scientific papers of this research problem. The creation and the use of this template is presented in the following paragraphs.

\subsubsection{Template creation}
After many comparisons, we found it necessary to provide a structure to organize the knowledge extracted from papers related to ontology learning. This structure allowed us to facilitate the organization of further relevant papers independently of the curator in a highly consistent knowledge graph.

\begin{table}[t]
	\caption{Table describing the classes of the template used to describe contributions of papers related to ontology learning}
	\centering
	\footnotesize
	\begin{tabular}{|p{3.5cm}|p{9cm}|}
	\hline
	\textbf{Class label}   &   \textbf{Example of instances}  \\
	\hline
	Knowledge source      & Text, databases, source code, etc. \\
	\hline
	Learning purpose & Constructing a new ontology, updating an existing ontology  \\
	\hline
	Application domain &   Medicine, Geography   \\
	\hline
	Learning data source & Java source code, XSD documents  \\
	\hline
	Has dataset &  300 source code files selected in the data source \\
	\hline
	Training corpus & 70\% of the dataset
	\\
	\hline
	Output format   & .txt, .owl, .json, .rdf, .xml
	\\
	\hline
	Input format 	& .txt, .XML, .png
	\\
	\hline
	Learning method &   Parser-based, Machine Learning-based, HMM, CNN
	\\
	\hline
	Learning tool   &   on-to-text, source2onto
	\\
	\hline
	Technologies	&   Java, Python, TensorFlow
	\\
	\hline
	Terms learning  &   Entities, shape, feature, aspects
	\\
	\hline
	Relationship	& Topological relation, Direction relation\\
	\hline
	Property	& DataProperties, ObjectProperties
	\\
	\hline
	Axiom   &   Transitive relation, reflexive relation
	\\
	\hline
	Rule	&  if(age<10)then children  
	\\
	\hline
	Evaluation  &   User evaluation, comparison to a gold standard
	\\
	\hline
	Knowledge assessment	& Empirical measure, human intervention, domain expert
	\\
	\hline
	\end{tabular}
	\label{tab:ontoLearnClasses}
\end{table}
To create the template, we used the properties we already added in the system for ontology learning from source code, database, UML models, etc. This template involves classes, properties (presented by the tables \ref{tab:ontoLearnClasses} and \ref{tab:properties}) applicable to a considerable number of papers related to ontology learning research papers. The comparisons elements that are created using this template are composed of instances of these classes and relations included in the template.

Each class is associated with a property that will appear in the comparison table as a comparison criteria in column property of the comparison table. In addition to these properties, other properties of basic data types are also added to the template. These properties are presented in the table \ref{tab:properties}.

\begin{table}[t]
	\centering
	\footnotesize
	\caption{Properties for comparing research contribution}
	\begin{tabular}{|p{4cm}|p{8.5cm}|}
	\hline
	\textbf{Property label}	&  \textbf{Description}  \\
	\hline
	Class learning	& True when the authors extracts classes from the data source \\
	\hline
	Instance learning 	& True when the authors extracts instances from the data source \\
	\hline
	Taxonomy learning 	& True when the authors extracts taxonomies of classes or properties from the data source \\
	\hline
	Class hierarchy learning   & True when the authors extracts class hierarchies from the data source \\
	\hline
	Validation tool	& Presents the technologies used to validate/develop the validation tool \\
	\hline
	Validation comments	& Any comments of the authors concerning the validation \\
	\hline
	Recall	& This is the recall of the learning tool \\
	\hline
	Precision	& This is the precision of the learning tool \\
	\hline
	F-measure 	& This is the F-measure of the learning tool \\
	\hline
	\end{tabular}
	\label{tab:properties}
\end{table}

\subsubsection{Using the template to create a new comparison}
The template presented in the section above was used to create 14 contributions. These contributions come from 2 papers related to ontology learning from images. To create these contributions, we identified the DOI of the papers found using Semantic Scholar. The DOI was entered using the "adding paper wizard" of ORKG. The system automatically extracts the papers metadata. Thereafter, knowledge was extracted manually and added to the system using the template. These contributions were finally used to create a comparison table. The graph visualization was used for the exploration of scientific contributions. It allowed us to realize that there was some confusion in our comparison. This confusion was corrected, the template and the comparison refined and new versions published.  A video presenting the curation of papers related to ontology learning from image was published by the ORKG team\footnote{\url{https://www.youtube.com/watch?v=EwfLJdPRr6o}}.

\subsubsection{Creation of smart review}
Once the informations are extracted from the papers related to ontology learning from images, these information were used to write a smart review. The goal of this review was to present and compare related work on ontology learning from image data.

\subsubsection{Collaborative work on literature review}
In this research we did not consider only our viewpoint during the creation of the template and comparison tables. We discussed with colleagues, other researchers using ORKG and the ORKG team to which we sent the links of these resources. This allowed us to refine them and create new versions. It should be noted that any fellow researchers can improve these resources with new information. For instance, if new literature is published, anyone can add a new contribution to the comparison table and publish a new version.

\section{An approach for knowledge acquisition from scientific papers}
\label{orkgApproach}

\begin{figure}
	\centering
	\caption{The description of the knowledge acquisition approach proposed in this paper}
	\includegraphics[angle=-90, scale=0.6]{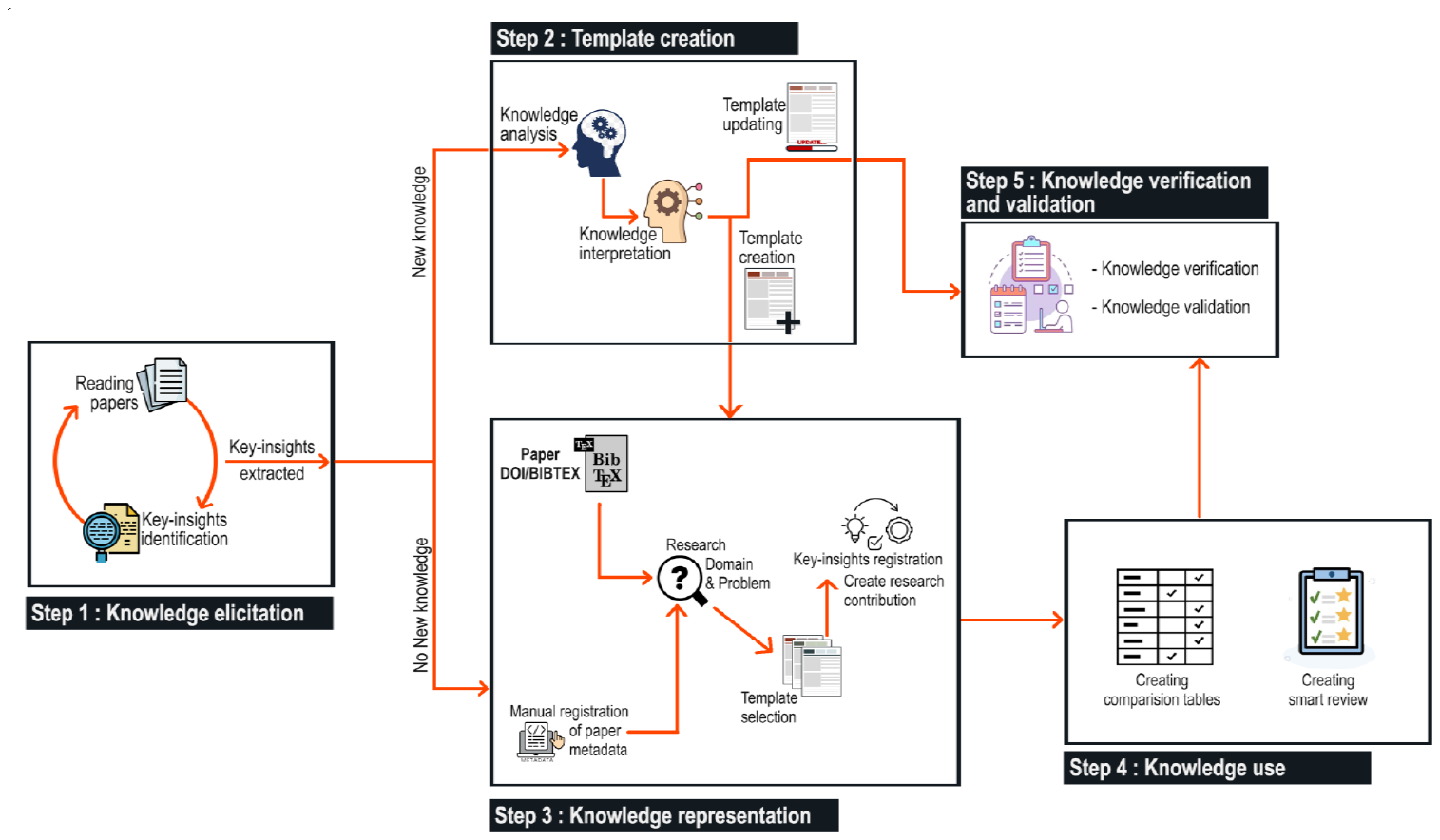}
	\label{fig:extraction_workflow}
\end{figure}

Acquiring knowledge from scientific papers from scratch is costly in time and resources. The approach we propose in this paper aims to reduce this cost during the knowledge acquisition process by allowing researchers to create structured repositories of scientific papers related to a research problem and/or a research domain. This approach is inspired by the use of ORKG in our research since three years to:
\begin{itemize}
	\item Organize and compare research contributions so as to build a large dataset of prior and up-to-date knowledge in our research domain;
	\item Organize research so as to facilitate the update and improvement with the contributions of fellow researchers working on the same research problem.
\end{itemize}
In effect, previously, to do a state of the art research, we were searching for relevant scientific papers on the Internet, reading these papers and summarizing them in text format and building comparison tables using LibreOffice Calc and Google Sheet. After the curation of ORKG in 2021, we got new insights on how to acquire and organize scientific literature. The latter is developed in this section as a computer-assisted knowledge acquisition approach from scientific papers (presented by Fig. \ref{fig:extraction_workflow}). It describes how knowledge can be extracted from research papers and stored in a knowledge graph in order to facilitate the access to key-insights hidden in research papers. It consists of six steps during which classes, properties and relations are extracted from scientific papers, and used to build a template. Thereafter, the template is used to represent contributions of the papers related to the same research problem. Finally, the contributions are used to build comparison tables, which themselves can be used to write a smart review. These steps are: Knowledge elicitation (Section \ref{approach:knowledgeElicitation}), Knowledge analysis and interpretation (Section \ref{approach:knowledgeAnalysis}), Templates creation (Section \ref{approach:templateCreation}), Knowledge representation (Section \ref{approach:knowledgeRepresentation}), Knowledge use (Section \ref{approach:knowledgeUse}) and Knowledge verification and validation (Section \ref{approach:knowledegVerifValidation}).

\subsection{Step 1: Knowledge elicitation}
\label{approach:knowledgeElicitation}
First and foremost, the researcher should determine the research domain that he wants to document. Thereafter he should identify the research problem related to this research domain. Once the research domain and the research problem are identified, these informations are used to search for relevant papers using search engines like Google search or search engines in digital research repositories like Semantic Scholar, Springer, Elsevier, IEEE, etc. For instance, in the domain of nutrition, a researcher may be interested in the food recommendation to people on diet. Thus, the following research question may be elicited: "How to recommend food to people in diet?" or "which techniques, methods and methodologies are used for food recommendation?". These research questions are used to search for scientific papers. Relevant papers related to this research domain and research problem are identified using many criteria which can be the title of the paper, the authors, references or citation analysis. References analysis can be used for instance to identify relevant papers to the research problem. During the selection of papers, the importance of a paper is defined as how close it is with the research domain and research problem. This task is done by reading the abstract or the full paper. Only papers that are too close to the research problem are selected. Once the research papers are found, some of them are selected for knowledge elicitation.

During the knowledge acquisition activity, the researcher should read the papers selected previously, identify and extract keywords, clauses, sentences, scientific claims, etc. Globally, all the information that is relevant to understand the paper is identified. This is an iterative process (see Fig. \ref{fig:extraction_workflow}) during which the researcher should be sure at the end that he has identified anything relevant. In early iterations of the cycle, the knowledge identified can refer to entities which are grouped, and will give classes. These classes will thus be put in relation with each other. At the end of this step all the relevant knowledge are extracted.

\begin{figure}
	\centering
	\caption{Representation of the triple: "HMM is used to extract information from source code"}
	\includegraphics[scale=0.15]{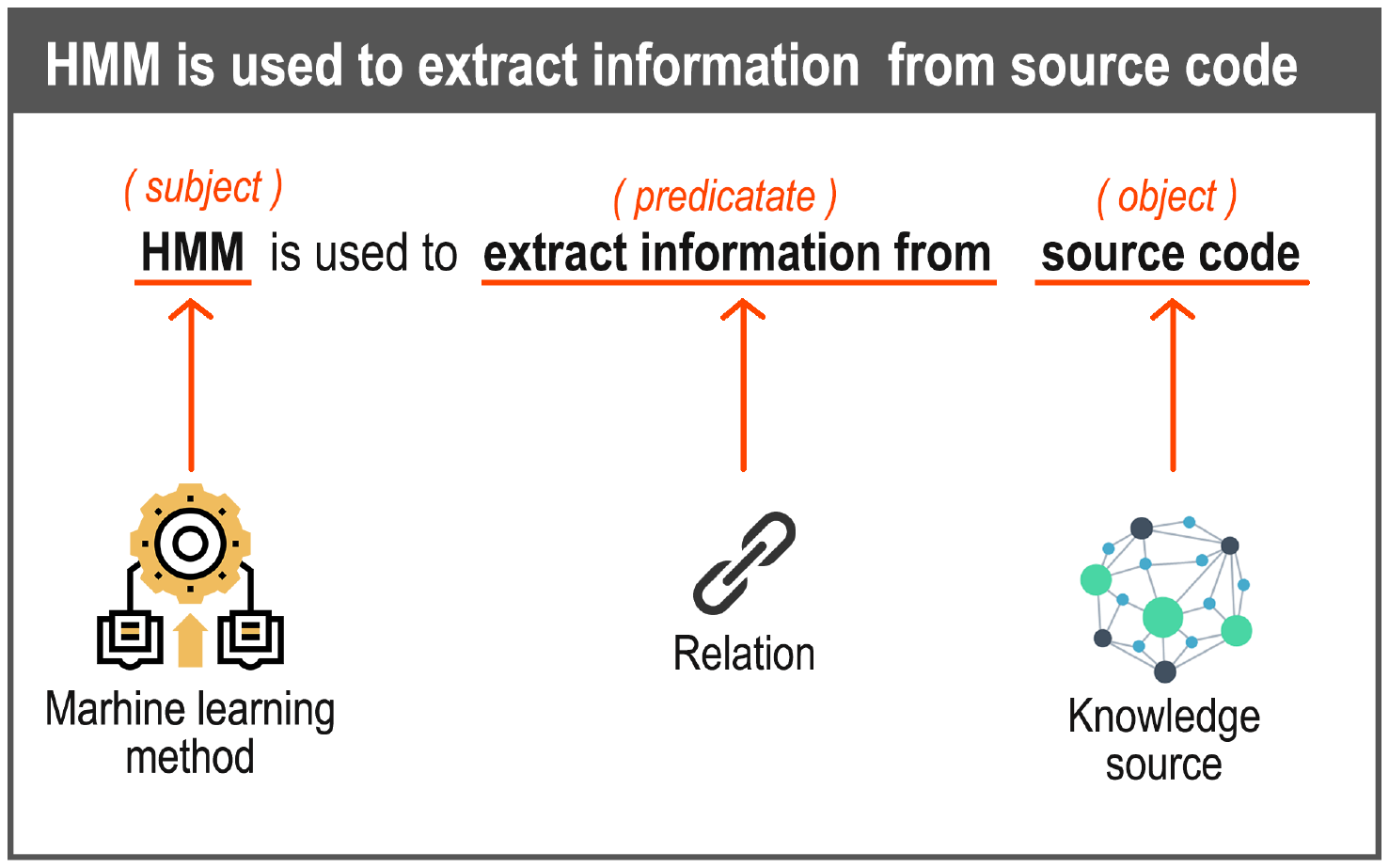}
	\label{fig:hmmTriple}
\end{figure}

The identification and the extraction process can be done by using handwritten notes, spreadsheet, or underline in order to highlight all the key-insights. Thereafter, each piece of information highlighted can be labeled with the type of knowledge it represents. For instance, if we highlighted "HMM is used to extract information from source code", then, we can label "HMM" as a Machine Learning method, "Source code" as a knowledge source and "extract information from" as a relation between the ML method and the knowledge source (Fig. \ref{fig:hmmTriple} presents this triple).

Given that within survey papers some overviews or summaries are often presented in (semi-)structured tabular format, the comparison criterion in these tables should be identified and extracted. These information could be extended with additional information extracted from this survey paper or other papers selected.

Globally, two kinds of information can be identified from the papers. We named them as keywords and keyphrases:
\begin{itemize}
	\item \textbf{Keywords:} keywords are words that are used to represent knowledge. For instance, if we consider the evaluation of ML techniques, we can identify the following keywords: "HMM", "Recall", "Precision", "Accuracy".
	\item \textbf{Keyphrases:} keyphrases are composed of a set of words that are used to represent a part of knowledge. For example, we have "Source code", "Wind power forecasting using time series".
\end{itemize}

\subsection{Step 2: Knowledge analysis and interpretation}
\label{approach:knowledgeAnalysis}
Knowledge analysis and interpretation consists of reviewing the elements extracted, identifying the key pieces of knowledge, providing a definition to each of these elements. Thereafter, these knowledge are assembled into related groups. Redundant informations are identified and only one term is selected. The definition of each keyword and keyphrases is provided.

Knowledge obtained after this task are classified into classes, relations, properties and instances. The terms in keywords and keyphrases are used to create the labels of these entities. During this task, the main challenge is to keep the keywords and keyphrases simple and descriptive.

\subsection{Step 3: Template creation}
\label{approach:templateCreation}
The classes, properties, relations and instances are used to create a template using ORKG template editor. This template is a conceptual model of papers dealing with the research domain and research problem addressed by its creator.

The template allows researchers to put key-insights hidden in research papers in a machine readable form. However, to be human readable, Classes, relations, properties and instances should have a definition in human readable form so that any human operator can use the template to register knowledge extracted from a paper. In order to create a consensus, the template link can be sent to researchers working in the research domain to have their point of view. To facilitate her/his amelioration, the author of the template can make the latter editable, so that other researchers can update.

\subsection{Step 4: Knowledge representation}
\label{approach:knowledgeRepresentation}
The knowledge representation step consists of using the template built in step 3 to annotate research papers related to the research domain and the research problem. Thus, the research contribution is machine and human-readable. Using the template, knowledge related to the research problem and research domain is continually refined and updated through additional knowledge from new scientific papers.

Globally, annotating a paper using ORKG and the template built in step 3 can be manual or (semi-)automatic. During the automatic process, the paper title, DOI or the BibText is entered in the add paper wizard. These metadata are used to fetch the paper and automatically extract other metadata. The next step of the process consists of selecting the research domain, defining the research problem and choosing the template to use in order to fill the other key-insights. Importing survey tables are also done (semi-automatically). Once the table is imported, the curator can correct information extracted and add additional key-insights. The manual process consists of adding the metadata and the key-insights manually.

Once ingested into ORKG, research contributions can be visualized as a semantic network. This graph can be used for the exploration of scientific contributions.

\subsection{Step 5: Knowledge use}
\label{approach:knowledgeUse}
Extracting knowledge from knowledge sources is not an end in itself. Once represented in a machine readable form, the knowledge acquired should be used. In our case, the knowledge acquired can be used to compare research papers and write smart reviews. In effect, the structured semantic representation of scientific knowledge in the KG makes it possible to automatically create literature comparisons. We are currently using these resources in our papers. One of these papers concerning "Food Composition Tables" is already published. The second one on "Food Information Engineering" was accepted at the AAAI conference.

\subsection{Step 6: Verification and validation}
\label{approach:knowledegVerifValidation}
The approach we present in this paper uses ORKG as an intelligent tool for assisting researchers in their work of organizing and comparing key-insights extracted from existing literature. Thus, in step 4 and 5, we show how it can be used to create research contributions and compare scientific papers. To ensure that the templates, contributions, comparisons tables and smart reviews contain the necessary elements and that these elements are well structured and presented, they should be verified and validated. To this end, any researcher who has an account on the ORKG platform can edit any comparison, template, modify and save (for templates) or publish a new version (for comparisons).

\section{Use cases}
\label{postIntervention}
Knowledge acquired during the intervention phase of the Action research methodology presented in Section \ref{researchMethodology:intervention}  were used to propose an approach using ORKG for knowledge acquisition from scientific papers (Section \ref{orkgApproach}). This section constitutes the Post-Intervention of the Action research during which this methodology is used in real world settings to solve related problems. This approach was used to curate over 200 papers corresponding to the "ontology learning", "epidemiological surveillance systems design and implementation", "food information engineering", "Tabular data to Knowledge Graph Matching", "Question Answering", and "information extraction from scientific papers'' research problems and, "Neuro-symbolic" domain\footnote{The overall work is freely available online at \url{https://orkg.org/u/ebdd4856-0ac9-4a65-a077-470fe2ca6826} and \url{https://orkg.org/u/aa79db4d-6762-4eb3-88fe-4db43405970c}}. From these research problems, we ingested over 800 contributions in the ORKG platform and we used these contributions to build over 100 comparisons tables. We used the template created during the curation of ORKG, and following steps 4 and 5 of the approach to create research contributions of papers related to "ontology learning from text" and "ontology learning from videos". The "knowledge use" step consists of creating comparison tables of "ontology learning from videos" and "ontology learning from text" research problems. The overall links to all the resources presented in this Section are given as additional materials. The rest of this section presents how this approach was applied step by step to curate 21 papers related to epidemiological surveillance systems (Section \ref{postIntervention:epiSurvDesignImplem}), how this approach is currently used to curate papers in the domain of food information engineering (Section \ref{postIntervention:foodInfoEngineering}) and how we used it to curate the papers used to write the related work of this research (Section \ref{postIntervention:KEScientificPapers}).

\subsection{Epidemiological surveillance systems}
\label{postIntervention:epiSurvDesignImplem}
Epidemiological surveillance systems enable the collection, analysis, and interpretation of data, together with the dissemination of these data to public health practitioners, clinicians, decision makers and the general population for preventing and controlling diseases \cite{PastPresentFuturPHSurv2012,CDCAdvanPHSur2017,MDAJiomekong2019}. It should support timely, efficient, flexible, scalable and interoperable data acquisition, analysis and dissemination. These informations are essential to the planning, implementation and evaluation of public health practices \cite{PastPresentFuturPHSurv2012,RalphFrerichs1991}. To design and implement epidemiological surveillance systems, it can be important to have an overview of existing systems. Thus, this section presents how the approach presented in section \ref{orkgApproach} is used to acquire knowledge from papers related to epidemiological surveillance and build a comparison table.

\subsubsection{Step 1: Knowledge elicitation}
To furnish relevant information to stakeholders, epidemiological surveillance systems should be designed and implemented so as to always correspond to the requirements. Thus, the current work is about the acquisition of key-insights on epidemiological surveillance design and implementation with the goal to identify approaches, techniques and tools that are used for epidemiological surveillance and to see the limits of existing systems.

Given that epidemiological surveillance systems are primarily concerned with the collection, analysis, interpretation and dissemination of information to different stakeholders, we choose to classify the papers related to "Epidemiological surveillance systems design and implementation" research problem in the domain of "information science".

Once the research problem and the domain are identified, we move to the searching and the selection of papers that will be used. The famous research repository "Semantic Scholar" were used to search for relevant research papers: (1) "epidemiological surveillance system" search string were entered in the search bar of Semantic Scholar; (2) "Computer Science were chosen" as the field of study.

We found 44600 papers. We used the papers titles, short abstract provided by Semantic Scholar and paper abstract provided by the authors to select relevant papers. Given the large number of papers retrieved, we decided to consider only the first page of results provided by Semantic Scholar. Thereafter, we went through the papers on the first page one by one, selecting those that seemed to be relevant given the research problem. Citations-based analysis was also used to search for relevant papers. This consists of identifying all the papers that are cited and that have been cited. Fortunately these papers are extracted and automatically presented by Semantic Scholar. From the papers identified as relevant, a total of 21 were selected randomly and downloaded.

Once the papers were selected, we divided these papers into two groups: a group of 4 papers for building the template and the rest. Knowledge was acquired from these 4 papers by the identification of important key terms from each paper selected. Thus, each paper was read line by line, and we identified from each of them key-insights that may be of interest to researchers. After the elicitation phase, an exact and complete transcript of the key-insights extracted were made.

\subsubsection{Step 2: Knowledge analysis and interpretation}
Knowledge that was saved in the transcript was reviewed and analyzed in order to identify key pieces of knowledge and their relationships that represent scientific information carried by the research paper. A deep analysis of elements extracted were used to identify classes, properties and relations as described in Section \ref{orkgApproach}. We were seeking the elements that are applicable to a considerable number of papers related to epidemiological surveillance systems.

\subsubsection{ Step 3: Template creation}
The main classes and properties identified during Step 2 were used to build a template of papers related to "epidemiological surveillance systems design and implementation\footnote{\url{https://www.orkg.org/orkg/template/R150089}}" research problem. This template is available online and can be improved by other researchers.

\subsubsection{Step 4: Knowledge representation}
During the knowledge representation step, the template built throughout the previous step was used to annotate the 4 papers used to build the template. Thereafter, knowledge were acquired from the rest of papers and ingested in ORKG using the template. In total, 21 papers were ingested in ORKG.

\subsubsection{Step 5: Knowledge use}
The contributions created using the template were used to build a comparison table\footnote{\url{https://www.orkg.org/orkg/comparison/R146851/}}. The latter compares papers related to the "epidemiological surveillance system design and implementation" research problem.

\subsubsection{Step 6: Knowledge verification and validation}
The discussion with a collaborator who is an epidemiologist allowed us to validate the template. They found that the template, the comparison table and the contributions constitute the elements that are helpful when putting in place an epidemiological surveillance system.

\subsection{Food Information Engineering}
\label{postIntervention:foodInfoEngineering}
Food information engineering involves the acquisition, the processing and the diffusion of up-to-date food information to different stakeholders. These informations are compiled from several data sources and used for a variety of purposes such as food recommendation, recipe substitution, food image recognition, etc. Many authors have proposed methodologies, methods and tools for the acquisition, the processing of food information, its storage, diffusion, etc. However, these contributions are scattered in many scientific papers on the Internet and are difficult to exploit. The second use case we chose consists of documenting the "food information engineering" research problem so as to provide to fellow researchers with methodologies, methods, tools, use cases, etc. It consists of documenting the following research question: "how food information is collected, processed, diffused and used?" To reply to this research question, several researches on the acquisition of food knowledge, its storage, querying and diffusion to different stakeholders are done worldwide. Our objective during this work is to document these solutions so as to provide to the research community with a body of knowledge that will help fellow researchers to reduce their research curve.

\subsubsection{Step 1: Knowledge elicitation}
Our goal during the Knowledge elicitation step was to identify several papers that can allow us to document the research question: "how food information is collected, processed, diffused and used?" We were having prior knowledge on the organization of food information using Food Composition Tables, Food Ontologies and Food Knowledge Graphs. Thus, we position food information engineering research problem in the Semantic Web research domain. Once the research problem to work on and the research domain was determined, we move to the searching of relevant papers. Our goal being to build comparison tables for the following research problems:
\begin{itemize}
	\item Food Composition Tables construction and description (5 papers processed, 4 comparisons tables built),
	\item Food Ontology construction, description and integration (27 papers processed, 4 comparison tables built and one smart review wrote),
	\item Food knowledge graph construction, description and integration (11 papers processed, 4 comparisons tables built and one smart review wrote),
	\item etc.
\end{itemize}

We used Google search to search for relevant papers from these subjects, using as keywords the title of the research problems. Only the first page (containing 10 research results) of the Google search platform was considered. In the case of "Food Ontologies" and "Food Knowledge Graphs", we used the most recent review published by Weiqing et al. \cite{FoodOntoKG_Weiqing_2022} to identify the research papers related to "Food Ontologies" and "Food Knowledge Graphs". Once retrieved, we choose some of these papers to identify elements that are comparable.

\subsubsection{Step 2: Knowledge analysis and interpretation}
As we did with epidemiological surveillance systems, knowledge that was identified from the papers downloaded and saved in transcript was reviewed and analyzed in order to identify classes, properties and relations. We used the comparison of "Food Ontologies" and "Food Knowledge Graphs" provided by Weiqing et al. \cite{FoodOntoKG_Weiqing_2022} to find additional properties. These tables were imported in the ORKG system\footnote{\url{https://orkg.org/comparison/R221127}}\footnote{\url{https://orkg.org/comparison/R217515/}}. In this particular case, we started by getting and using all properties used to compare papers in the review paper.

During the analysis of papers, we found that many authors were providing Question Answering Systems over Food KG. Thus, we decided to document this research problem\footnote{\url{https://orkg.org/comparison/R239314}}\footnote{\url{https://orkg.org/comparison/R269002}}.

\subsubsection{ Step 3: Templates creation}
Once the papers were selected, they were used to build new templates (8 templates) and update existing templates (two templates were updated). For instance, the template of "Ontology description" created during the Intervention phase (Section \ref{researchMethodology:intervention}) was updated with new properties, another template created by Natalia Chichkova, an ORKG user for the description of KG was also updated. The following examples of templates were created by zero and are currently used: food composition tables, Question Answering systems and Question Answering benchmark.

\subsubsection{Step 4: Knowledge representation}
During the knowledge representation step, the template built throughout the previous step was used to annotate all the papers downloaded by considering different research problems. Currently, more than 120 papers related to the domain of "Food Information Engineering" are ingested in the ORKG platform. It should be noted that this is an ongoing work and we want at its end to provide to the research community with a systematic literature review of "food information engineering" research problems.

\subsubsection{Step 5: Knowledge use}
The contributions created using the template were used to build over 26 comparison tables. The comparison table of "food composition table" research papers allowed us to realize that "food composition tables" change over time and unfortunately, the database did not change. On the other hand, the supports used to distribute these data are sparse on the Internet in different formats. We also realize that up-to-date data can be found in scientific papers. Thus, we build a large scale and up-to-date food composition tables that is currently annotated using Wikidata.

\subsubsection{Step 6: Knowledge verification and validation}
Knowledge validation consists of presenting this work in challenges and conferences. Our work on "Food Composition Table" was accepted in SemTab challenge \footnote{\url{https://sem-tab-challenge.github.io/2022/}} organized by International Semantic Web Conference 2022\footnote{\url{https://iswc2022.semanticweb.org/}}. The overall work on food information engineering was accepted at "New Faculty Highlights" AAAI-23\footnote{\url{https://aaai.org/Conferences/AAAI-23/new-faculty-highlights-cfp/}} conference program. We are currently adding more papers in order to maintain a state of the art of papers in the domain of "food information engineering".

\subsection{Knowledge extraction from scientific papers}
\label{postIntervention:KEScientificPapers}
The process of literature review starts from the searching of scientific papers from the huge amount of existing ones to the analysis of the paper content and the extraction of key-insights from them. Given the large amount of scientific papers in all domains, this process is laborious, time consuming and cumbersome. To reduce the burden of work, knowledge extraction from scientific papers is of great interest to researchers. During the last years, this research problem has interested many researchers and methodologies, methods and tools have been proposed. Our goal during the third use case was to identify the different types of knowledge that are extracted from scientific papers and to document datasets, methodologies, models and tools used for extracting these knowledge.

\subsubsection{Step 1: Knowledge elicitation}
Given that the research problem we are documenting is "knowledge extraction", we classified this research problem in the Semantic Web domain. As we did with the two previous use cases, our goal during this step was to identify several papers that can cover the research question we want to document. By using the search keyword: "knowledge extraction from scientific paper" on Google Search engine, we found a great survey \cite{surveyIEScienPaperZara2018}. This is a 60 pages survey of datasets, methodologies, methods and tools that are used to extract different types of knowledge from scientific papers. It is organized in two main sections: (1) Metadata extraction, (2) Key-insights extraction. Each section describes the different types of knowledge that are extracted, the methods that are used to extract each type of knowledge and the evaluation of each method. We found this survey interesting for knowledge elicitation.

The survey paper was read line by line in order to identify elements that are comparable. The comparisons tables provided by the authors were great resources for the identification of key-insights. Thus, we combine the knowledge extracted from these tables with the knowledge extracted from the full body text to obtain a set of key-insights candidates.

\subsubsection{Step 2: Knowledge analysis and interpretation}
The key-insights identified from the tables and the text were analyzed one by one in order to select the ones that can be considered as relevant. The duplicates were also identified and deleted.

\subsubsection{ Step 3: Templates creation}
Knowledge identified during the previous step were converted into properties, classes and relations. Thereafter, these classes, properties and relations were used to create templates. We found it necessary the create the following templates:
\begin{itemize}
	\item Template for metadata dataset\footnote{\url{https://orkg.org/template/R277000}}: this template is used to describe the content of each metadata dataset.
	\item Templates for Key-Insight\footnote{\url{https://orkg.org/template/R279223}}\footnote{\url{https://orkg.org/template/R280533}}: two types of datasets describing key-insights were found: sentence-level key-insight and phrase-level key-insights. These templates are used to describe these datasets.
	\item Template of metadata system\footnote{\url{https://orkg.org/template/R280212}}: this is used to describe the different systems that are used for extracting the metadata from the scientific article.
	\item Template of key-insight system\footnote{\url{https://orkg.org/template/R280523}}: this is used to describe the different systems that are used for extracting key-insights from scientific papers.
\end{itemize}
In addition to these templates, we reused a template\footnote{\url{https://orkg.org/template/R259041}} that we created during the work on "food information engineering" for evaluating each extraction system. We also used a template\footnote{\url{https://orkg.org/template/R166722}}, created by  Jennifer D'Souza for the description of existing tools that are proposed for knowledge extraction from scientific papers.

\subsubsection{Step 4: Knowledge representation}
During the knowledge representation step, the template built throughout the previous step was used to annotate papers related to "information extraction from scientific papers".

\subsubsection{Step 5: Knowledge use}
Currently, more than 50 papers related to "information extraction from scientific papers" are being ingested in ORKG. These papers are used to document the "information extraction from scientific papers" research problem. From these papers, more than 50 research contributions were extracted and used to build 11 comparison tables. These resources were used to write the related work of this research (see Section \ref{relatedWork}). 

\subsubsection{Step 6: Knowledge verification and validation}
The templates and the contributions provided in this research will be evaluated by the reviewers of this paper. On the other hand, these resources can be evaluated, validated and improved by any researcher working on knowledge extraction from scientific papers.

\section{Related work}
\label{relatedWork}
As presented in the previous sections, scientific knowledge can be grouped into two categories: metadata and key-insights \cite{surveyIEScienPaperZara2018, KEReviewAbdul2020}.

During the last years, many researchers have contributed in the domain of metadata extraction from research papers. Zara et al \cite{surveyIEScienPaperZara2018} and Abdul et al. \cite{KEReviewAbdul2020} present a great state-of-the-art on this subject. These works show that manual processing is generally used for scientific papers annotations in order to build datasets. Thereafter, these datasets are used to train models that will further be used for metadata extraction. The models that are used for metadata extraction are rule-based, machine learning-based and Natural Language Processing-based. Rule-based models use text features and layouts to define instructions that specify how to extract desired information from scientific papers. On the other hand, methods such as Hidden Markov Models (HMM), Conditional Random Fields (CRF), Support Vector Machines (SVM), Neural Networks are also proposed for metadata extraction from scientific papers. The approaches proposed for metadata extraction are very powerful. The evaluation of the most powerful ones show the performances reaching 95\% of F-measure.

Key-insights acquisition consists of reading the scientific paper, identifying relevant knowledge and organizing them or building models for their automatic extraction. In the rest of this section, we present the different types of key-insights in Section \ref{relatedWork:keyInsights}, existing key-insights datasets in Section \ref{relatedWork:datasets}, methods for key-insights extraction in Section \ref{relatedWork:methods}, and tools for key-insights extraction in Section \ref{relatedWork:tools}.

\subsection{Key-insights}
\label{relatedWork:keyInsights}
Key-insights are presented in scientific papers in the form of text, figures and tables. The semi-structured organization of knowledge in tabular data allows us to easily extract key-insights from tables stored in scientific papers. For instance, Food Composition Tables can be extracted in scientific papers for accessing food that people are eating and their nutritive values \cite{Jiomekong2022largeFCT}. However, key-insights hidden in text are more difficult to identify and extract because it is difficult to guess the valuable information enclosed within a research paper text that can be beneficial for each researcher. Zara et al. \cite{surveyIEScienPaperZara2018} classified key-insights hidden in the paper text into sentence-level key-insights, phrase-level key-insights, and relation \cite{surveyIEScienPaperZara2018}:
\begin{itemize}
	\item \textbf{Sentence-level key-insights:} these are predefined knowledge, in the form of keywords and key-phrases and hidden in the text of an article. For instance, "method", "problem", "objective", "result", etc. are included in almost all scientific papers.
	\item \textbf{Phrase-level key-insights:} These are phrases carrying potential information that are useful to researchers. For instance, "tool or library", "measures and measurements", "language resource product", "location", etc.
	\item \textbf{Relation:} relation can express application of a technique to solve a problem, results generated against various evaluation measures, etc. Phrase-level key-insights can be extended to extract relations because in many cases, relations are expressed between entities.
\end{itemize}
Key-insights acquisition from scientific papers can be done manually or automatically. We presented in Sections \ref{orkg}, \ref{orkgCuration}, \ref{orkgApproach} and \ref{postIntervention} how ORKG can be used as a computer-assistant tool for semi-automatic acquisition of knowledge from scientific papers. To build models for automatic acquisition (or extraction) of key-insights from scientific papers, there is a need for annotated datasets. In the next section, we present related work on key-insights datasets.

\subsection{Datasets}
\label{relatedWork:datasets}
Based on the different types of key-insight that can be extracted from scientific papers, the datasets for extracting these knowledge can be classified as Sentence-level key-insights and Phrase-level key-insight datasets.
\begin{itemize}
	\item \textbf{Sentence-level key-insights datasets:} These datasets contain scientific articles in which sentences are classified based on insights they carry. We gathered the different properties that can be used to compare sentence-level key-insights and we built an ORKG template. Thereafter, this template was used to compare Sentence-level key-insights datasets published in scientific literature.

	\item \textbf{Phrase-level key-insight datasets:} these datasets contain scientific papers in which phrases are annotated with entities corresponding to potential key-insights they may carry. The datasets for phrase-level key-insight extraction are difficult to build and scarce. As we did with sentence-level key-insights, we built an ORKG template of phrase-level key-insights and we used this template to compare phrase-level key-insights datasets.
\end{itemize}
The comparison of phrase-level and sentence-level key-insights shows that the majority of existing datasets belong to the domain of medical science. On the other hand, these datasets are mainly based on the extraction of knowledge from the abstract only \cite{surveyIEScienPaperZara2018}.

\subsection{Acquisition methods}
\label{relatedWork:methods}
Acquiring knowledge from scientific papers can be manual or automatic. Automatic knowledge acquisition relies on rules, Machine Learning, Deep Learning and Natural Language Processing techniques for automatic identification and extraction of key-insights. Based on the datasets presented in Section \ref{relatedWork:datasets}, Zara et al. \cite{surveyIEScienPaperZara2018} classified these methods as sentence-level key-insights and phrase-level key-insights methods.
\begin{itemize}
	\item \textbf{Sentence-level key-insights extraction:} methods for Sentence-level key-insights extraction are focused on the classification of sentences in predefined categories based on insights they carry.
	\item \textbf{Phrase-level key-insights extraction:} methods for Phrase-level key-insights extraction are focused on the extraction of phrases carrying potential information.
\end{itemize}
To extract sentence-level and phrase-level key-insights from scientific papers, rules, ML, DL and NLP techniques have been proposed. The main techniques proposed are Bayesian classifier, Conditional Random Field, Support Vector Machine, Hidden Markov Models. To compare research work on this subject, we built a template and we used this template to compare several methods for sentence-level and phrase-level key-insights extraction. These methods are not as powerful as metadata extraction methods. Very little works show methods that the F-measure reaches 85\% for the extraction of each key-insights.

\subsection{Tools for knowledge acquisition}
\label{relatedWork:tools}
Key-insights acquired from scientific papers are generally grouped into research contributions to make them comparable with other resources. To this end, hand-written notes can be used to organize and build comparison tables and figures. Tools used for knowledge acquisition from scientific papers aim to facilitate this work and make it less laborious, time consuming and cumbersome. They can be classified as computer-assisted tools, tools for automatic extraction of key-insights, digital research repositories and social tagging and bookmarking platforms.

\subsubsection{Computer-assisted tools}
Computer-assisted tools aim to help researchers to organize key-insights extracted from scientific papers. Spreadsheets software such as Microsoft Excel, Libreofficel Cal or Google spreadsheets are generally used to organize, store and compare research contributions from several research papers. The main advantages of these software is that the data can be stored and reused whenever needed. It is also easier to build graphics with the data. However, these data are not harmonized, isolated in researcher computers or storage and difficult to merge with other research data. Thus, two researchers will make the same effort to extract the same knowledge from a set of scientific papers. These efforts can be saved if the knowledge is organized in a computer-assisted software such as ORKG. In a recent work, Allard et al. \cite{ComparingReConAllard2019} present a workflow designed to compare research contributions in ORKG. This paper shows the process to add a paper and the key-insights of this paper in ORKG. However, it did not provide a complete methodology from the knowledge elicitation phase (using template to create a conceptual model of the domain) to the knowledge use phase.

\subsubsection{Digital research repositories}
Digital research repositories aim at providing researchers with basic filters to ease the search of scientific papers while querying through millions of research papers. To this end, metadata informations are used to provide various searching facilities \cite{surveyIEScienPaperZara2018}. On the other hand, key-insights are used to augment keywords and provide short abstracts (e.g., Semantic Scholar) to guide researchers to identify relevant papers to his research problem.

\subsubsection{Social tagging and bookmarking platforms}
Social tagging and bookmarking platforms (e.g. CiteULike, Bibsonomy, Delicious) are online services for serving scientific communities \cite{KEReviewAbdul2020}. The users of these tools can annotate the research articles, bookmark the preferences, etc. This allows them to possess their references or a web page with their own defined tags or keywords. But this does not allow researchers to compare research contributions identified from several research papers.

Even if the knowledge of some Digital research repositories and Social tagging and bookmarking platforms are organized in knowledge graphs (e.g., Springer Nature SciGraph\footnote{\url{https://www.springernature.com/gp/researchers/scigraph}}, Microsoft Academic \cite{WangMicrosoftGraph2020}), these tools does not permit to researchers to structure key-insights hidden so to help other researchers to update with more papers and insights.

\section{Summary and conclusion}
\label{conclusion}
Acquiring knowledge from scientific papers from scratch is costly in time and resources. Thus, we propose in this paper an approach using Open Research Knowledge Graph as a computer-assistant tool for knowledge acquisition from scientific papers. It consists of five steps:
\begin{itemize}
	\item Knowledge elicitation consists of determining the domain and the research problem to document. Using these information, to search for relevant scientific papers and extract elements that one wants to compare.
	\item Knowledge analysis and interpretation consist of analyzing the pertinence of the elements extracted during knowledge elicitation and the deletion of duplicates.
	\item Template creation consists of using the elements obtained after the knowledge analysis and interpretation to build a template that will be used further to organize key-insights extracted and research contributions.
	\item Knowledge representation consists of using existing templates to structure knowledge extracted in a knowledge graph.
	\item Knowledge use consists of comparing research contributions in comparison tables, and using them to write reviews of the domain.
	\item Verification and validation consists of the validation of the templates, the contributions, the comparisons of research contributions and the reviews by fellow researchers.
\end{itemize}
This approach is currently used to document the "ontology learning", "epidemiological surveillance systems design and implementation", "food information engineering", "Tabular data to Knowledge Graph Matching", "Question Answering", and "information extraction from scientific papers" research problems and the "Neuro-symbolic AI" domain. Thus, more than 200 papers are ingested in ORKG. From these papers, more than 800 contributions are documented and these contributions are used to build over 100 comparison tables. At the end of this work, we found that ORKG is a valuable tool that can reduce the working curve of state-of-the-art research.

\section*{Acknowledgement}
We are grateful to the Open Research Knowledge Graph team for their following during the curation of ORKG. Our great thanks also goes also to all the curators. Their remarks and questions were very helpful in this work.


\bibliographystyle{elsarticle-num-names}
\bibliography{paper_submitted}

\begin{thebibliography}{25}
\expandafter\ifx\csname natexlab\endcsname\relax\def\natexlab#1{#1}\fi
\providecommand{\url}[1]{\texttt{#1}}
\providecommand{\href}[2]{#2}
\providecommand{\path}[1]{#1}
\providecommand{\DOIprefix}{doi:}
\providecommand{\ArXivprefix}{arXiv:}
\providecommand{\URLprefix}{URL: }
\providecommand{\Pubmedprefix}{pmid:}
\providecommand{\doi}[1]{\href{http://dx.doi.org/#1}{\path{#1}}}
\providecommand{\Pubmed}[1]{\href{pmid:#1}{\path{#1}}}
\providecommand{\bibinfo}[2]{#2}
\ifx\xfnm\relax \def\xfnm[#1]{\unskip,\space#1}\fi
\bibitem[{Jayaram and Sangeeta(2017)}]{IEResearchPapersJayaram2017}
\bibinfo{author}{K.~Jayaram}, \bibinfo{author}{K.~Sangeeta},
\newblock \bibinfo{title}{A review: Information extraction techniques from
  research papers},
\newblock in: \bibinfo{booktitle}{2017 International Conference on Innovative
  Mechanisms for Industry Applications (ICIMIA)}, \bibinfo{year}{2017}, pp.
  \bibinfo{pages}{56--59}. \DOIprefix\doi{10.1109/ICIMIA.2017.7975532}.
\bibitem[{Shah et~al.(2003)Shah, Perez{-}Iratxeta, Bork, and
  Andrade}]{IEArticleParantu2003}
\bibinfo{author}{P.~K. Shah}, \bibinfo{author}{C.~Perez{-}Iratxeta},
  \bibinfo{author}{P.~Bork}, \bibinfo{author}{M.~A. Andrade},
\newblock \bibinfo{title}{Information extraction from full text scientific
  articles: Where are the keywords?},
\newblock \bibinfo{journal}{{BMC} Bioinform.} \bibinfo{volume}{4}
  (\bibinfo{year}{2003}) \bibinfo{pages}{20}. \URLprefix
  \url{https://doi.org/10.1186/1471-2105-4-20}.
  \DOIprefix\doi{10.1186/1471-2105-4-20}.
\bibitem[{Auer et~al.(2020)Auer, Oelen, Haris, Stocker, D’Souza,
  Eddine~Farfar, Vogt, Prinz, Wiens, and Jaradeh}]{ImprovingAccessAuer2020}
\bibinfo{author}{S.~Auer}, \bibinfo{author}{A.~Oelen},
  \bibinfo{author}{M.~Haris}, \bibinfo{author}{M.~Stocker},
  \bibinfo{author}{J.~D’Souza}, \bibinfo{author}{K.~Eddine~Farfar},
  \bibinfo{author}{L.~Vogt}, \bibinfo{author}{M.~Prinz},
  \bibinfo{author}{V.~Wiens}, \bibinfo{author}{M.~Y. Jaradeh},
\newblock \bibinfo{title}{Improving access to scientific literature with
  knowledge graphs},
\newblock \bibinfo{journal}{BIBLIOTHEK – Forschung und Praxis}
  (\bibinfo{year}{2020}). \DOIprefix\doi{http://dx.doi.org/10.18452/22049}.
\bibitem[{Nasar et~al.(2018)Nasar, Jaffry, and
  Malik}]{surveyIEScienPaperZara2018}
\bibinfo{author}{Z.~Nasar}, \bibinfo{author}{S.~W. Jaffry},
  \bibinfo{author}{M.~K. Malik},
\newblock \bibinfo{title}{Information extraction from scientific articles: a
  survey},
\newblock \bibinfo{journal}{Scientometrics} \bibinfo{volume}{117}
  (\bibinfo{year}{2018}) \bibinfo{pages}{1931--1990}. \URLprefix
  \url{https://doi.org/10.1007/s11192-018-2921-5}.
  \DOIprefix\doi{10.1007/s11192-018-2921-5}.
\bibitem[{Shahid et~al.(2020)Shahid, Afzal, Abdar, Basiri, Zhou, Yen, and
  Chang}]{KEReviewAbdul2020}
\bibinfo{author}{A.~Shahid}, \bibinfo{author}{M.~T. Afzal},
  \bibinfo{author}{M.~Abdar}, \bibinfo{author}{M.~E. Basiri},
  \bibinfo{author}{X.~Zhou}, \bibinfo{author}{N.~Y. Yen},
  \bibinfo{author}{J.~Chang},
\newblock \bibinfo{title}{Insights into relevant knowledge extraction
  techniques: a comprehensive review},
\newblock \bibinfo{journal}{J. Supercomput.} \bibinfo{volume}{76}
  (\bibinfo{year}{2020}) \bibinfo{pages}{1695--1733}. \URLprefix
  \url{https://doi.org/10.1007/s11227-019-03009-y}.
  \DOIprefix\doi{10.1007/s11227-019-03009-y}.
\bibitem[{Adnan and Akbar(2019)}]{LimitIETechKiran2019}
\bibinfo{author}{K.~Adnan}, \bibinfo{author}{R.~Akbar},
\newblock \bibinfo{title}{Limitations of information extraction methods and
  techniques for heterogeneous unstructured big data},
\newblock \bibinfo{journal}{International Journal of Engineering Business
  Management} \bibinfo{volume}{11} (\bibinfo{year}{2019})
  \bibinfo{pages}{1847979019890771}. \DOIprefix\doi{10.1177/1847979019890771}.
\bibitem[{Acheson and Purves(2021)}]{IEGISPapersAcheson2021}
\bibinfo{author}{E.~Acheson}, \bibinfo{author}{R.~S. Purves},
\newblock \bibinfo{title}{Extracting and modeling geographic information from
  scientific articles},
\newblock \bibinfo{journal}{PLOS ONE} \bibinfo{volume}{16}
  (\bibinfo{year}{2021}) \bibinfo{pages}{1--19}. \URLprefix
  \url{https://doi.org/10.1371/journal.pone.0244918}.
  \DOIprefix\doi{10.1371/journal.pone.0244918}.
\bibitem[{Caroline~Hayes(1996)}]{KAGuideHayes1996}
\bibinfo{author}{D.~G. Caroline~Hayes, Michael~Fu}, \bibinfo{title}{Some
  Guidelines for Knowledge Acquisition Strategies}, \bibinfo{type}{Technical
  Report}, AAAI Technical Report SS-96-02, \bibinfo{year}{1996}. \URLprefix
  \url{https://www.aaai.org/Papers/Symposia/Spring/1996/SS-96-02/SS96-02-008.pdf}.
\bibitem[{Gasevic et~al.(2009)Gasevic, Djuric, and Devedzic}]{Djuric2005}
\bibinfo{author}{D.~Gasevic}, \bibinfo{author}{D.~Djuric},
  \bibinfo{author}{V.~Devedzic}, \bibinfo{title}{Model Driven Engineering and
  Ontology Development}, \bibinfo{edition}{2nd} ed.,
  \bibinfo{publisher}{Springer Publishing Company, Incorporated},
  \bibinfo{year}{2009}.
\bibitem[{Karl et~al.(2013)Karl, Herrick, Unnasch, Gillan, Ellis, Lutters, and
  Martin}]{DiscoveringEcologicallyKarl2013}
\bibinfo{author}{J.~W. Karl}, \bibinfo{author}{J.~E. Herrick},
  \bibinfo{author}{R.~S. Unnasch}, \bibinfo{author}{J.~K. Gillan},
  \bibinfo{author}{E.~C. Ellis}, \bibinfo{author}{W.~G. Lutters},
  \bibinfo{author}{L.~J. Martin},
\newblock \bibinfo{title}{{Discovering Ecologically Relevant Knowledge from
  Published Studies through Geosemantic Searching}},
\newblock \bibinfo{journal}{BioScience} \bibinfo{volume}{63}
  (\bibinfo{year}{2013}) \bibinfo{pages}{674--682}.
  \DOIprefix\doi{10.1525/bio.2013.63.8.10}.
\bibitem[{Margulies et~al.(2016)Margulies, Magliocca, Schmill, and
  Ellis}]{AmbiguousGeographiesJared2016}
\bibinfo{author}{J.~D. Margulies}, \bibinfo{author}{N.~R. Magliocca},
  \bibinfo{author}{M.~D. Schmill}, \bibinfo{author}{E.~C. Ellis},
\newblock \bibinfo{title}{Ambiguous geographies: Connecting case study
  knowledge with global change science},
\newblock \bibinfo{journal}{Annals of the American Association of Geographers}
  \bibinfo{volume}{106} (\bibinfo{year}{2016}) \bibinfo{pages}{572--596}.
  \DOIprefix\doi{10.1080/24694452.2016.1142857}.
\bibitem[{Kamel(2008)}]{KAMagdi2007}
\bibinfo{author}{M.~N. Kamel},
\newblock \bibinfo{title}{Knowledge acquisition},
\newblock in: \bibinfo{editor}{B.~W. Wah} (Ed.), \bibinfo{booktitle}{Wiley
  Encyclopedia of Computer Science and Engineering}, \bibinfo{publisher}{John
  Wiley {\&} Sons, Inc.}, \bibinfo{year}{2008}.
  \DOIprefix\doi{10.1002/9780470050118.ecse205}.
\bibitem[{Oelen et~al.(2019)Oelen, Jaradeh, Farfar, Stocker, and
  Auer}]{ComparingReConAllard2019}
\bibinfo{author}{A.~Oelen}, \bibinfo{author}{M.~Y. Jaradeh},
  \bibinfo{author}{K.~E. Farfar}, \bibinfo{author}{M.~Stocker},
  \bibinfo{author}{S.~Auer},
\newblock \bibinfo{title}{Comparing research contributions in a scholarly
  knowledge graph},
\newblock in: \bibinfo{editor}{D.~Garijo}, \bibinfo{editor}{M.~Markovic},
  \bibinfo{editor}{P.~Groth}, \bibinfo{editor}{I.~Santana{-}P{\'{e}}rez},
  \bibinfo{editor}{K.~Belhajjame} (Eds.), \bibinfo{booktitle}{Proceedings of
  the Third International Workshop on Capturing Scientific Knowledge co-located
  with the 10th International Conference on Knowledge Capture {(K-CAP} 2019),
  Marina del Rey, California , November 19th, 2019}, volume
  \bibinfo{volume}{2526} of \textit{\bibinfo{series}{{CEUR} Workshop
  Proceedings}}, \bibinfo{year}{2019}, pp. \bibinfo{pages}{21--26}.
\bibitem[{et~al.(2020)}]{empiricalStandSoftEnRalph2020}
\bibinfo{author}{R.~et~al.}, \bibinfo{title}{Empirical standards for software
  engineering research}, \bibinfo{year}{2020}. \URLprefix
  \url{https://arxiv.org/abs/2010.03525}.
  \DOIprefix\doi{10.48550/ARXIV.2010.03525}.
\bibitem[{Azanzi et~al.(2019)Azanzi, Camara, and Tchuente}]{AzanziCT2019}
\bibinfo{author}{F.~J. Azanzi}, \bibinfo{author}{G.~Camara},
  \bibinfo{author}{M.~Tchuente},
\newblock \bibinfo{title}{Extracting ontological knowledge from java source
  code using hidden markov models},
\newblock \bibinfo{journal}{Open Computer Science} \bibinfo{volume}{9}
  (\bibinfo{year}{2019}) \bibinfo{pages}{181--199}. \URLprefix
  \url{https://doi.org/10.1515/comp-2019-0013}.
  \DOIprefix\doi{10.1515/comp-2019-0013}.
\bibitem[{Konys(2018)}]{KONYS20182194}
\bibinfo{author}{A.~Konys},
\newblock \bibinfo{title}{Knowledge systematization for ontology learning
  methods},
\newblock in: \bibinfo{booktitle}{Knowledge-Based and Intelligent Information
  {\&} Engineering Systems: Proceedings of the 22nd International Conference
  KES-2018, Belgrade, Serbia, 3-5 September 2018.}, \bibinfo{year}{2018}, pp.
  \bibinfo{pages}{2194--2207}. \URLprefix
  \url{https://doi.org/10.1016/j.procs.2018.07.229}.
  \DOIprefix\doi{10.1016/j.procs.2018.07.229}.
\bibitem[{Shamsfard and
  Abdollahzadeh~Barforoush(2003)}]{ShamsfardMehrnoush2003}
\bibinfo{author}{M.~Shamsfard}, \bibinfo{author}{A.~Abdollahzadeh~Barforoush},
\newblock \bibinfo{title}{The state of the art in ontology learning: A
  framework for comparison},
\newblock \bibinfo{journal}{Knowl. Eng. Rev.} \bibinfo{volume}{18}
  (\bibinfo{year}{2003}) \bibinfo{pages}{293--316}.
\bibitem[{Zhou(2007)}]{zhou2007ontology}
\bibinfo{author}{L.~Zhou},
\newblock \bibinfo{title}{Ontology learning: state of the art and open issues},
\newblock \bibinfo{journal}{Information Technology and Management}
  \bibinfo{volume}{8} (\bibinfo{year}{2007}) \bibinfo{pages}{241--252}.
  \URLprefix \url{https://doi.org/10.1007/s10799-007-0019-5}.
  \DOIprefix\doi{10.1007/s10799-007-0019-5}.
\bibitem[{C~K~Choi(2012)}]{PastPresentFuturPHSurv2012}
\bibinfo{author}{B.~C~K~Choi},
\newblock \bibinfo{title}{The past, present, and future of public health
  surveillance},
\newblock \bibinfo{journal}{Scientifica} \bibinfo{volume}{2012}
  (\bibinfo{year}{2012}) \bibinfo{pages}{875253}.
\bibitem[{Richards et~al.(2017)Richards, Iademarco, Atkinson, Pinner, Yoon,
  Kenzie, Lee, Qualters, and Frieden}]{CDCAdvanPHSur2017}
\bibinfo{author}{C.~L. Richards}, \bibinfo{author}{M.~F. Iademarco},
  \bibinfo{author}{D.~Atkinson}, \bibinfo{author}{R.~W. Pinner},
  \bibinfo{author}{P.~Yoon}, \bibinfo{author}{W.~R.~M. Kenzie},
  \bibinfo{author}{B.~Lee}, \bibinfo{author}{J.~R. Qualters},
  \bibinfo{author}{T.~R. Frieden},
\newblock \bibinfo{title}{Advances in public health surveillance and
  information dissemination at the centers for disease control and prevention},
\newblock \bibinfo{journal}{Public Health Reports} \bibinfo{volume}{132}
  (\bibinfo{year}{2017}) \bibinfo{pages}{403--410}.
\bibitem[{Jiomekong and Camara(2019)}]{MDAJiomekong2019}
\bibinfo{author}{A.~Jiomekong}, \bibinfo{author}{G.~Camara},
\newblock \bibinfo{title}{Model-driven architecture based software development
  for epidemiological surveillance systems},
\newblock \bibinfo{journal}{Studies in health technology and informatics}
  \bibinfo{volume}{264} (\bibinfo{year}{2019}) \bibinfo{pages}{531—535}.
  \URLprefix \url{https://doi.org/10.3233/SHTI190279}.
  \DOIprefix\doi{10.3233/shti190279}.
\bibitem[{Frerichs(1991)}]{RalphFrerichs1991}
\bibinfo{author}{R.~R. Frerichs},
\newblock \bibinfo{title}{Epidemiologic surveillance in developing countries},
\newblock \bibinfo{journal}{Annual Review Public Health} \bibinfo{volume}{12}
  (\bibinfo{year}{1991}) \bibinfo{pages}{257}.
\bibitem[{Min et~al.(2022)Min, Liu, Xu, and Jiang}]{FoodOntoKG_Weiqing_2022}
\bibinfo{author}{W.~Min}, \bibinfo{author}{C.~Liu}, \bibinfo{author}{L.~Xu},
  \bibinfo{author}{S.~Jiang},
\newblock \bibinfo{title}{Applications of knowledge graphs for food science and
  industry},
\newblock \bibinfo{journal}{Patterns} \bibinfo{volume}{3}
  (\bibinfo{year}{2022}) \bibinfo{pages}{100484}.
  \DOIprefix\doi{https://doi.org/10.1016/j.patter.2022.100484}.
\bibitem[{Jiomekong et~al.(2022)Jiomekong, Etoga, Foko, Tsague, Folefac, Kana,
  Sow, and Camara}]{Jiomekong2022largeFCT}
\bibinfo{author}{A.~Jiomekong}, \bibinfo{author}{C.~Etoga},
  \bibinfo{author}{B.~Foko}, \bibinfo{author}{V.~Tsague},
  \bibinfo{author}{M.~Folefac}, \bibinfo{author}{S.~Kana},
  \bibinfo{author}{M.~M. Sow}, \bibinfo{author}{G.~Camara},
\newblock \bibinfo{title}{A large scale corpus of food composition tables},
\newblock \bibinfo{journal}{Semantic Web Challenge on Tabular Data to Knowledge
  Graph Matching (SemTab), CEUR-WS. org}  (\bibinfo{year}{2022}).
\bibitem[{Wang et~al.(2020)Wang, Shen, Huang, Wu, Dong, and
  Kanakia}]{WangMicrosoftGraph2020}
\bibinfo{author}{K.~Wang}, \bibinfo{author}{Z.~Shen},
  \bibinfo{author}{C.~Huang}, \bibinfo{author}{C.-H. Wu},
  \bibinfo{author}{Y.~Dong}, \bibinfo{author}{A.~Kanakia},
\newblock \bibinfo{title}{{Microsoft Academic Graph: When experts are not
  enough}},
\newblock \bibinfo{journal}{Quantitative Science Studies} \bibinfo{volume}{1}
  (\bibinfo{year}{2020}) \bibinfo{pages}{396--413}.
  \DOIprefix\doi{10.1162/qss_a_00021}.

\end{thebibliography}




\end{document}